%% file: mw_bar.tex
\documentclass[useAMS,fleqn,usenatbib]{mnras}

\usepackage[T1]{fontenc}
\usepackage{ae,aecompl}

\usepackage{ marvosym }

\usepackage{amsmath}
\usepackage{amsopn}
\usepackage[british]{babel}
\usepackage[varg]{txfonts}
\usepackage{bm}
\usepackage{biblio} 
\usepackage{natbib}
\usepackage{color}
\usepackage{tcolorbox}
\usepackage{graphicx}

\usepackage{microtype}

\input{defs_mnras}

\newcommand{\myemail}{tepper@physics.usyd.edu.au}
\newcommand{\ramses}{{\sc ramses}}
\newcommand{\galaxia}{{\sc galaxia}}
\newcommand{\agama}{{\sc agama}}
\newcommand{\gaia}{{\em Gaia}}

\title[ A barred MW Surrogate ]{ A barred Milky Way surrogate from an N-body simulation}

\author[Tepper-Garc\'\i{}a et al.]{%
Thor Tepper-Garc\'\i{}a,$^{1,2,3}$\thanks{Contact e-mail: \href{mailto:\myemail}{\myemail}}
Joss Bland-Hawthorn,$^{1,3}$
Eugene Vasiliev,$^4$
E.~Athanassoula,$^5$
\newauthor
Ortwin Gerhard,$^6$
Alice Quillen,$^7$
Paul McMillan,$^8$
Ken Freeman,$^9$
Geraint F.~Lewis,$^{1}$
\newauthor
Romain Teyssier,$^{9}$
Sanjib Sharma,$^{1,3}$
Michael R.~Hayden$^{1,3}$ and
Sven Buder$^{3,10}$
\\
$^1$ Sydney Institute for Astronomy, School of Physics, The University of Sydney, NSW 2006, Australia\\
$^2$ Centre for Integrated Sustainability Analysis, School of Physics, The University of Sydney, NSW 2006, Australia\\
$^3$ Centre of Excellence for All Sky Astrophysics in Three Dimensions (ASTRO-3D), Australia\\
$^4$ Institute of Astronomy, Madingley road, Cambridge, CB3 0HA, UK\\
$^5$ Aix Marseille Univ, CNRS, CNES, LAM, Marseille, France \\
$^6$ Max Planck Institute for Extraterrestrial Physics, 85741 Garching, Germany \\
$^7$ Department of Physics and Astronomy, University of Rochester, Rochester, NY 14627, USA \\
$^8$ Lund Observatory, Department of Astronomy and Theoretical Physics, Lund University, Box 43, SE-22100, Lund, Sweden\\
$^{9}$ Princeton University, Department of Astrophysical Sciences, 4 Ivy Lane, Princeton, New Jersey, 08544, United States of America\\
$^{10}$ Research School of Astronomy \& Astrophysics, Australian National University, Canberra, ACT 2611, Australia \\
}

\date{Accepted ---. Received ---; in original form ---}

\pubyear{\date{year}}

\begin{document}
\label{firstpage}
\pagerange{\pageref{firstpage}--\pageref{lastpage}}
\maketitle

\pdfminorversion=5
\begin{abstract}

We present an N-body model for the barred Milky Way (MW) galaxy that reproduces many of its properties, including the overall mass distribution, the disc kinematics, and the properties of the central bar.
Our high-resolution (N $\sim 10^8$ particles) simulation, performed with the \ramses\ code, starts from an axisymmetric non-equilibrium configuration constructed within the \agama\ framework. This is a self-consistent dynamical model of the MW defined by the best available parameters for the dark matter halo, the stellar disc and the bulge. 

For the known (stellar$+$gas) disc mass ($4.5\pm 1.5 \times 10^{10}$ \Msun) and disc mass fraction at $R \approx 2.2 R_d$ ($f_{\rm d} \approx 0.3-0.6$), the low mass limit does not yield a bar in a Hubble time. The high mass limit adopted here produces a box/peanut bar within about 2 Gyr with the correct mass (M$_{\rm bar} \sim 10^{10}$ \Msun), size (R$_{\rm bar} \approx 5$ kpc) and peak pattern speed ($\Omega_{\rm bar} \approx 40-45$ km s$^{-1}$ kpc$^{-1}$).
In agreement with earlier work, the bar formation timescale scales inversely with $f_{\rm d}$ (i.e. $\log T_{\rm bar}({\rm Gyr}) \approx 0.60/f_{\rm d} - 0.83$, $1\lesssim f_{\rm d} \lesssim 0.3$). The disc radial heating is strong, but, in contrast to earlier claims, we find that disc vertical heating outside of the box/peanut (b/p) bulge structure is negligible.

The synthetic barred MW exhibits long-term stability, except for the slow decline ($\dot{\Omega}_{\rm bar} \approx -2$ km s$^{-1}$ kpc$^{-1}$ Gyr$^{-1}$) of the bar pattern speed, consistent with recent estimates. If our model is indicative of the Milky Way, we estimate that the bar first emerged $3-4$ Gyr ago.




\end{abstract}


\begin{keywords}
Galaxy: general - Galaxy: evolution - methods: numerical
\end{keywords}

%


\section{Introduction} \label{s:intro}

There is a major international investment in all-sky stellar surveys, as exemplified
by ESA \gaia\ satellite mission, and by many ground-based photometric and multi-object spectroscopic surveys.
The goals of these surveys vary but they are united in their quest to shed light on the structure and evolution of the Milky Way \citep[for comprehensive reviews, see][]{freeman2002,Rix2013,Helmi2020}. In recent times, even our most basic understanding of the Milky Way has been called into question. Specifically, what aspects of the Milky Way are defined by long-term equilibrium and stability (if any), and which features are entirely transitory \citep[][chapter 6]{Binney2008}.

With the aim of providing a consistent framework for comparing data and models, we attempt to build an N-body ``surrogate'' for the Milky Way defined by the best parameters we have for the major components \citep{blandhawthorn2016}. Many researchers have considered the prospect of
a consistent framework with a view to interpreting
the \gaia\ data releases. One approach is to identify Milky Way analogues within cosmological (zoom) simulations.
But perhaps the most serious disadvantage of this approach is the impossibility to find galaxies that match the MW in every aspect of interest, e.g. a disc with the appropriate mass and scalelength, featuring a central bar embedded in a DM halo of the appropriate mass \citep[e.g.][]{Scannapieco2012,gra17h,ElBadry2018,Fragkoudi2020,Santistevan20,age21l}. These typically have low to modest resolution. 

Another approach is the tailored N-body simulation developed by \cite{Sellwood1984} with an accretion history designed to produce a synthetic analogue of the Milky Way \citep[e.g.][]{Berrier2016}. Ultimately, cosmologically motivated models are highly desirable because they recognize that our Milky Way has built up over 13 billion years through a process of accretion and evolution, and it experienced secular evolution due to the influence of the bar, spiral arms and molecular clouds. These are typically far less expensive to run than zoomed universe simulations such that many thousands of realizations can be carried out \citep{aum17a}.

An intermediate solution is to include gas and its physics in the simulation, while using a simplified way of describing the environment. This of course leads to considerably fewer simulations than the thousands possible with the Aumer et 
al. approach, but can still be sufficient to reach useful conclusions. 

For example, \citet[][see also \citealt{Peschken2017}]{Athanassoula2016} ran several hundreds of such simulations, with between five and thirty million particles each and a linear resolution of 25 to 50 pc. One of their most interesting aspects is that they provide accurate information on the age of all stars, and, provided they have been coupled to a chemical evolution code, can provide very useful chemodynamic information \citep{Athanassoula2017}. This was performed for about a dozen amongst the few hundreds of simulations of the Athanassoula et al. sample. Such simulation results can be compared with results from recent spectroscopic surveys. Indeed, \citet{Athanassoula2017} reproduced well the links between metallicity and kinematics found previously by the Abundances and Radial velocity Galactic Origins Survey (ARGOS) spectroscopic survey \citep{Freeman2013,Ness2013}.

Arguably the best Galactic bulge/bar realisation has come from made-to-measure (M2M) model building \citep{Portail2017} but the solutions to date are restricted to the bulge-bar-inner disc region. Their M2M technique is a sophisticated adaption of existing N-body models where orbits are perturbed and their weights adapted to arrive at an optimal configuration specified by effective potential parameters and observational data constraints. This method recognizes inherently that N-body models to date do not provide surrogates for the Milky Way without additional, and to some extent arbitrary, guidance by the modeller.
The idea of orbit weighting to arrive at a consistent gravitational potential-density pair goes back to \cite{Schwarzschild1979} who used
it to build spheroidal galaxies.


Our scheme is to construct an isolated, self-consistent galaxy with the right parameters using the \agama\ initialization code \citep{Vasiliev2019}; this is a controlled experiment where the components can be chosen to produce {\it inter alia} either an axisymmetric disc system with long-term stability, or an emergent barred galaxy with slow evolution. The prior work that most resembles our approach is \citet{Fujii2019} who employ the GalactICs code \citep{Widrow2008} to set up their GPU-based N-body simulations. In our view,
the Galactic parameters are sufficiently well understood for this to be worth the effort \citep{McMillan2017}, but, as \cite{blandhawthorn2016} make clear, there is no guarantee that the outcome will exhibit long-term stability given the uncertainties in the observational constraints. 
Despite the use of independent initial set-up codes and N-body solvers, our results share many similarities, as well as a few differences with \citet{Fujii2019}, which we discuss
throughout the text.
The \agama\ code is sufficiently new and versatile that we include a summary of the procedures specific to our work in Sec. 3 described below.

An obvious limitation embedded in our philosophy, and by inference in \citet{Fujii2019}'s, is the assumption that the Galaxy we observe today exhibits long-term stability of {\it any} kind, or that its evolution (e.g. growth through accretion) is sufficiently slow at the present epoch. We evolve our model for $\sim 4$ Gyr, in contrast to \citet{Fujii2019} who evolve theirs over 10 Gyr. As illustrated in \citet[][their figure 1]{blandhawthorn2016}, the Milky Way's mass has barely changed over a lookback time of 4 Gyr but has grown by a factor of $2-3$ over the last 10 Gyr. Yet, at present
the Milky Way is clearly undergoing a strong interaction with the Magellanic system and the Sgr dwarf galaxy \citep[e.g.][]{Laporte2018}, and it is also undeniable that it had strong interactions in the past, as evidenced by the recent discovery of a strongly radially anisotropic population in the stellar halo produced by an ancient merger \citep{Belokurov2018,Helmi2018}. It is worth stressing that the extent to which these events obviate the assumption of an isolated model is unclear, because these perturbing satellites excite modes that dampen over a few rotation periods in the inner disc \citep{Widrow2014}.

On the other hand, there are good reasons for believing that the Galactic bar has been in place for at least a few billion years \citep{Eskridge2000,Cole2002,James2016,Baba2020}. In support of this, in realistic simulations, the peak amplitude (Fourier mode $A_2$) from the bar instability can be attained after several Gyr \citep[e.g.][]{Athanassoula2005b,Fujii2018}.

Thus, we seek to produce a self-consistent dynamical model where the central bar retains its main properties, or exhibits slow evolution on Gyr timescales. Thus, we consider 4 Gyr (roughly 20 complete orbits at the Solar Circle) to be an appropriate timescale for an isolated model of the Milky Way.

Our work improves over earlier N-body models of the Galaxy in the following aspects: (1) we produce a barred spiral galaxy with properties that {\it closely resemble} the Milky Way today, including the bar's mass, size, corotation radius, pattern speed and slow-down rate; (2) the particle number of our N-body model is significantly higher compared to earlier similar studies \citep[with the notable exception of][]{Fujii2019}; (3) we make the full simulation data publicly available.

Our Milky Way surrogate model provides an important framework within which to pursue a series of experiments, in particular: (i) to study the emergence of spiral density waves and resonances before and after the bar forms;
(ii) to interpret the observed radial velocity streaming in the local stellar disc \citep{eil19a};
(iii) to explore secular processes like radial migration in different stellar populations \citep{Sellwood2002};
(iv) to separate the role of the bar and the dynamical impulse when the surrogate interacts with an interloper or, equivalently, to identify the dominate excitation modes before and after the disc transit;
(v) in principle, it can be inserted into a (static) mock survey like the Besan\c{c}on model \citep{robin2012} or the \galaxia\ code \citep{sharma2011}.\footnote{Both codes are axisymmetric and insert a bulge potential into a disc potential by excavating a hole in the inner disc.}

In Section 2, we explain our approach to building a minimalist Milky Way model. In Section 3, we describe the set up and initial conditions for our models
as prescribed by \agama. The simulations are described  in Section 4, before presenting the results in Section 5.  A key omission in our model is the cool gas phase that is to be included in our next incarnation. While there have been discussions of {\em long-term} equilibrium in galaxy models with embedded cool gas \citep[e.g.][]{Rodionov2011,Deg2019}, these have yet to be demonstrated; this is the subject of a follow-up paper.

\section{Minimalist Galaxy Models} \label{s:models}

We construct our surrogate Milky Way after carefully reviewing what has been learnt about disc stability over the past 60 years \citep[q.v.][]{Binney2008}.  There is a reasonable understanding of what stabilizes and destabilizes a rotating disc residing within a (live) dark matter halo, what are the triggers for bar and spiral arm formation and transience, and so on \citep[e.g.][]{Athanassoula2003,Fujii2018}. Including a live, instead of a rigid halo, is particularly important for the bar formation and evolution, as shown by \citet{Athanassoula2002} and by \citet{Sellwood2016}.

We focus here on the simplest type of galactic model, which consists of three {\em live} components: 1) a DM host halo; 2) a stellar bulge; and 3) a stellar disc, henceforth referred to as an HBD system or HBD galaxy. These are the three dominant non-gaseous components of the Milky Way by mass \citep{blandhawthorn2016}. Thus, we neglect the effects of the environment (e.g. accretion events) on the evolution of the Galaxy.

As discussed in \citet[][Sec. 4.2]{blandhawthorn2016},
for many years, the Galactic bulge was considered to be a structure built through mergers early
in the formation of the Milky Way, now called a `classical' bulge. We have assumed a classical bulge in our model, in line with Milky Way-like N-body models published to date.
But it is important to note that the assumption of a classical bulge may not be valid. The observed bar/bulge may be entirely a property of the bar instability in the disc \citep{Shen2010}.
Our inclusion of a classical bulge leads to a higher mass within the bar region in our model and a higher rotation curve in the centre (see Secs. ~\ref{s:match} and ~\ref{s:bar4}).

\citet{Fujii2018} have conducted an extensive study based on numerical simulations, which underlined two well-known and remarkable results: 1) {\em Any} N-body HBD system is bar-unstable ultimately; 2) The timescale for the onset of the bar instability is mainly governed by $f_{\rm d}$, the ratio of disc mass to total galaxy mass within the radius at which the rotation curve roughly peaks, where
\begin{equation}
\label{e:fd}
    f_{\rm d}= \left(\frac{V_{\rm \phi, disc}(R_s)}{V_{\rm \phi, tot}(R_s)}\right)_{R_s=2.2 R_d}^2 \, .
\end{equation}
Here, $R=R_d$ is the exponential disc scalelength and $V_\phi(R)$ is the circular velocity at a radius $R$. The higher the value of $f_{\rm d}$, the shorter the timescale for the onset of the bar instability. For this ratio evaluated at $R_s=2.2 R_d$, these authors establish that the dividing line occurs at $f_{\rm d}\approx 0.35$, in good agreement with \citet{val17a}, in the sense that smaller ratios lead to bar formation timescales that exceed a Hubble time. 


\subsection{Axisymmetric stable and $\mbox{m=2}$ unstable models} \label{s:m2}


In successive papers, \cite{Fujii2018} and \cite{Fujii2019} modelled a Milky Way-like disc with of order 10$^7$ and 10$^9$ particles, respectively. Using a simulation size that sits midway between these two limits,
we confirm their result for the low $f_{\rm d}$ limit; a bar is not observed over
the lifetime of the synthetic disc \citep{BlandHawthorn2021}.
This is an important test because these authors also demonstrate that simulation size (i.e. particle number) and bar formation timescale are coupled in the `low N' limit. Other effects, in many cases of less importance, influencing the precise value of this threshold, and the timescale for the onset of the bar instability, include the kinematic temperature of the disc (quantified by \citealt{too64a}'s $Q$ parameter), and the bulge-to-disc mass ratio.

Thus, \citet{Fujii2018} essentially provide a recipe for the construction of models that can be considered as bar `stable' ($f_{\rm d} < 0.35$) or bar unstable ($f_{\rm d} \gtrsim 0.35$) over a Hubble time. Based on this insight, we have designed two HBD galaxies: 1) A galaxy that is stable for at least 4.3 Gyr \citep[discussed in detail in][]{BlandHawthorn2021}; 2) a galaxy that it is unstable to the formation of a bar within a reasonable time scale ($\approx 2$ Gyr). We refer to these models as the `smooth' galaxy and the `barred' galaxy from here on.

To arrive at our (ultimately successful) barred galaxy model, we tune the mass of each component (DM halo, bulge, stellar disc) such that the disc-to-total mass ratio given by eq. ~\eqref{e:fd} peaks well above the minimum threshold, while keeping the mass of all components within their observed ranges. In fact, the properties of both of our models are consistent with the Galaxy (e.g. rotation curve, component masses, etc.; cf. \citealt{blandhawthorn2016}, and our Fig.~\ref{f:ics1} ); for further details on our smooth model, we refer the reader to \citet{BlandHawthorn2021}.



\section{Initial conditions} \label{s:ics}

The initial conditions (particle positions and particle velocities) of our N-body galaxy models are created with the Action-based GAlaxy Modelling Architechture software package \citep[\agama{};][]{Vasiliev2019}. In the following, we describe the procedure in detail, as it may prove instructive to others wishing to follow our approach.

\subsection{Theoretical background}

The evolution of a collisionless N-body system with a sufficiently large number of particles is governed by the collisionless Boltzmann equation and its solution, the `one-particle' distribution function, $f(\bm{r},\bm{v},t)$. In a steady state, \mbox{$f(\bm{r},\bm{v},t) \equiv f(\bm{r},\bm{v},t=0)$}, and $f$ is a function of the integrals of motion (constants) only \citep{jea15a}.

A convenient set of integrals of motion are the action variables, which in the case of axisymmetric potentials, are the radial action $J_R$, the vertical action $J_z$ and azimuthal action $J_{\phi}$ (equivalent to the $z$-component of the angular momentum vector).
In this case, \mbox{$f = f(\bm{J}) \equiv f(J_R,J_z,J_\phi)$}. This implies the existence of a mapping \mbox{$(\bm{r},\bm{v}) \mapsto \bm{J}$} that depends on the total potential of the system $\psi_{tot}$, succinctly expressed as\footnote{The mapping also includes the three angles variables, which we ignore henceforth because the integral over these variables simply yields a factor $(2 \pi)^3$ which is absorbed into the normalisation of the DF.} $\bm{J}\left[\bm{r},\bm{v} ~\vert ~\psi_{tot} \right]$.

The distribution function (DF) provides a statistical description of the dynamical state of the system, and other properties can be derived from it. For example, for an appropriate normalisation, the mass density $\rho_c$ of a component (e.g. stellar disc) described by a DF $f_c$ is given by
\begin{equation} \label{e:dens}
    \rho_c(\bm{r}) = \int \!\! f_c(\bm{J}[\bm{r},\bm{v}]) ~d\bm{v} \,.
\end{equation}
%

\subsection{AGAMA initialisation} \label{s:agama}

Within the context of \agama, a fully self-consistent, equilibrium galaxy model with a specified number of components (e.g. halo, bulge, disc as we do here) is defined by the following:
\begin{enumerate}
    \item The total potential of the system, $\psi_{tot}$, determined via Poisson's equation and the superposition of all the components' density
    \item The mapping $\bm{J}\left[\bm{r},\bm{v} ~\vert ~\psi_{tot} \right]$. \agama\ makes use of the so-called `St\"ackel fudge' \citep{bin12u}, appropriate for axisymmetric potentials
    \item The DF of each component expressed in terms of actions, $f_c(\bm{J})$
    \item The density of each component calculated from its DF (Eq.~\ref{e:dens})
\end{enumerate}
These translate into a coupled system of non-linear equations that can be solved iteratively. The `self-consistent-modelling' module - one of many within the \agama\ library - adopts the following prescription:
\begin{itemize}
    \item Specify the initial density profile $\rho_c$ of each galaxy component
    \item Specify the DF, $f_c(\bm{J})$, for each component; $f_c$ remains fixed during subsequent iterations
    \item Calculate the potential of the system via $\nabla \psi_{tot} = 4 \pi \sum_c \rho_c$
    \item Determine the mapping $\bm{J}\left[\bm{r},\bm{v} ~\vert ~\psi_{tot} \right]$ and recompute the new density of each component $\rho'_c$ via Eq.~\eqref{e:dens}
    \item Repeat the last two steps until `convergence' (as defined below)
\end{itemize}
Convergence during the iterative process is guaranteed by the adiabatic invariant nature of actions, together with the use of $f(\bm{J})$. This is best judged by the maximum absolute change of the total potential across iterations, and ideally it should be on the order of a percent or less. 
Once a converged model has been attained, an N-body representation of the system is created by drawing a specified number $N$ of random samples (point-like masses or `particles') for each component directly from phase space $(\bm{r}, \bm{v})$ by evaluating $f(\bm{J})$ using the mapping $\bm{J}\left[\bm{r},\bm{v} ~\vert ~\psi_{tot} \right]$ together with the final, self-consistent potential. To this end, \agama\ implements an efficient multidimensional sampling rejection algorithm.
Thus the construction of an N-body model with \agama\ yields a fully self-consistent \mbox{`potential - density - velocity distribution'} triplet.

In terms of the procedure outlined above, all that is required from the user's perspective are the first two steps. \agama\ takes care of the rest, including the sampling of the density and velocity distributions (for a user specified $N$). The choice of density profiles, and of the DFs and their parameters should be motivated by physical considerations and/or constrained by observations, as in our case (see next section). 

We now describe our \agama\ initial setup in detail. For all models, {\em `initial'} refers to the state of the model prior to the iterative process performed with the help of \agama; {\em `final'} refers to the state of the model upon achieving convergence during the iterative process. Later when we discuss our \ramses\ simulations (Sec. \ref{s:sim}), we will refer to an {\em `unevolved'} model -- the state of the model at the beginning of the simulation ($T = 0$ Gyr) -- and to an {\em `evolved'} model, meaning the state of model during the simulation at some $T > 0$ Gyr. Thus, the \agama\ {\em final} model and the \ramses\ {\em unevolved} model are identical.

\subsection{Initial density profiles and DFs} \label{s:prof}

In the following we describe in detail our choice of parameters to generate initial conditions for our synthetic barred galaxy. The setup to create our smooth galaxy using \agama\ has been described elsewhere \citep{BlandHawthorn2021}.

\subsubsection{Density profiles} \label{s:dens}

The initial density of the DM halo and of the stellar bulge is described by the following function,\footnote{The spheroidal profile in \agama\ depends on far more parameters than included in Eq.\eqref{e:sph}, but we list only those that are relevant to our modelling; the omitted parameters all take their \agama\ default values. We refer the reader to the \agama\ documentation \citep{Vasiliev2018} for details. \label{fon:doc} }
\begin{equation} \label{e:sph}
    \rho_s(r) = \rho_0 \left( \frac{r}{a} \right)^{-\gamma} \left[ 1 + \left( \frac{r}{a} \right) \right]^{\gamma - \beta} \times \exp{\left[ - \left( \frac{r}{r_c} \right)^2 \right]} \, ,
\end{equation}
that represents a double power-law (`spheroidal') profile with a taper. Here, $r$ is the spherical radius; the other parameters have the following meaning:
\begin{itemize}
    \item $\rho_0$ := density normalisation
    \item $a$ := scale radius
    \item $r_c$ := outer cut-off radius
    \item $\beta$ := outer power-law slope
    \item $\gamma$ := inner power-law slope
\end{itemize}
%

\begin{table}
\begin{center}
\caption{Parameter values that define the initial density profiles of the spheroidal components in our \agama\ setup. See text around Eq.~\ref{e:sph} for the definition of the parameters.}
\label{t:sph}
\begin{tabular}{lcccccc}
Component & $\rho_0$ & $a$ & $\beta$ & $\gamma$ & $r_c$ & $M_{tot}$ \\
 & (\Msun ~\kpc$^{-3}$) & (kpc) & & & (kpc) & ($10^{10}$ \Msun) \\
\hline
DM halo     & $9\times10^9$ & 19 & 3 & 1 & 250 & 117 \\
Stellar bulge       & -- & 0.6 & 4 & 1 & 2 & 1.3
\end{tabular}
\end{center}
\begin{list}{}{}
\item Notes. $M_{tot}$ and $\rho_0$ are not both independent parameters for any type of component. For the DM halo, we specify the  density normalisation; the total mass depends on it, and is listed only for convenience. For the bulge, we specify the total mass rather than the density normalisation. 
\end{list}
\end{table}

Our choice of parameter values for the spheroidal components (DM halo, stellar bulge) are listed in Tab.~\ref{t:sph}. These choices yield a DM halo that roughly follows a NFW profile \citep{nav97a}, and a stellar bulge with a profile close to a Hernquist profile \citep{her90a}.
While the choice of profile and the precise value of the structural parameters are somewhat arbitrary, the total mass and the extension of the spheroidal components are in good agreement with the corresponding estimates for the dark halo and the stellar bulge of the Milky Way \citep{blandhawthorn2016}.

It is worth emphasising that the parameter values listed in Tab.~\ref{t:sph} are the initial parameter values that determine the initial potential of the iterative process required for the construction of a self-consistent N-body model (see previous section). The {\em final} values at convergence, i.e. the end of the iterative procedure -- in particular the structural parameters -- may differ, albeit only slightly, from their initial values. The same is true for the stellar disc.
The upshot is that the model density and corresponding potential are not given in a closed, analytic form, but represented as two-dimensional (2D) spline functions in appropriately scaled, cylindrical coordinates.\\ 

The initial density profile for the stellar disc is given by
\begin{equation} \label{e:disc}
    \rho_d(R,z) = \frac{\Sigma_0}{4 | h |} \exp{ \left[ - \frac{R}{R_d} \right]} \times \sech^2{\left| \frac{z}{2 h} \right|} \, ,
\end{equation}
where $R$ is the cylindrical radius, $z$ is the vertical distance from the plane, and the other parameters have the following meaning:
\begin{itemize}
    \item $\Sigma_0$ := surface density normalisation
    \item $R_d$ := scalelength
    \item $h$ := scaleheight
\end{itemize}

Our choice of parameter values that define the initial density profile of the stellar disc are given in Tab.~\ref{t:disc}.

\begin{table}
\begin{center}
\caption{Parameter values that define the  initial density profile of the stellar disc. See text around Eq.~\ref{e:disc} for the definition of the parameters. }
\label{t:disc}
\begin{tabular}{lccccc}
Component & $\Sigma_0$ & $R_d$ & $h$ & $R_c$ & $M_{tot}$ \\
 & (\Msun ~\kpc$^{-2}$) & (kpc) & (kpc) & (kpc) & ($10^{10}$ \Msun) \\
\hline
Stellar disc       & $1.1\times10^9$ & 2.5 & -0.3 & 0 & 4.3
\end{tabular}
\end{center}
\begin{list}{}{}
\item Notes: The convention in \agama\ is that $h < 0$ defines a vertical $\sech^2$ profile.
\end{list}
\end{table}

\begin{table*}
\begin{center}
\caption{ Type of distribution function and corresponding parameter values used in our \agama\ setup to initialise the stellar disc.
}
\label{t:dfdisc}
\begin{tabular}{lcccccc}
Type & $\tilde{\Sigma}_0$ & $\tilde{R}_d$ & $\tilde{h}$ & $\sigma_{R,0}$ &  $R_{\sigma_R}$ & $J_{min}$ \\
 & ($10^9$ \Msun\ \kpc$^{-2}$) & (kpc) & (kpc) & (\kms) & (kpc) & (\kms ~\pkpc) \\
\hline
Quasi-Isothermal & $1.1$ & 2.5 & 0.3 & 70 & 5 & 40 \\
\hline
\end{tabular}
\end{center}
\begin{list}{}{}
\item Notes: The parameter $J_{min}$ is set to roughly the angular momentum of a circular orbit at the scale radius of the bulge (see Tab.~\ref{t:sph}), to deter the stellar disc from featuring a `hole' in the centre.
\end{list}
\end{table*}

\subsubsection{Distribution functions} \label{s:df}

By virtue of \citet{jea15a}'s theorem, {\em any} analytic function of the actions $(J_R, J_z, J_\phi)$ is a solution to the Boltzmann equation, and thus a suitable DF. Therefore, one has the freedom to choose functional forms that are appropriate for the system under study, assuming one can calculate the actions. \agama\ implements a number of algorithms to calculate actions from coordinates and velocities, notably the `St\"ackel fudge'. The library also provides a number of DFs to choose from. From the user's perspective, the most important consideration when making a choice is whether a component is `spheroidal' or `discy'.\footnote{ For more details on the different types of DF, their advantages and disadvantages, we refer the reader to \citet[][his section 5]{Vasiliev2019}. }

For the DM halo and the bulge we choose a DF of the type `Quasi-Spherical', well-suited to model `spheroidal' components. For the stellar disc, we use an improved variant of the `Quasi-Isothermal' DF, which is appropriate for a `discy' component.

A Quasi-Spherical DF is constructed directly from the density distribution using either the Eddington inversion formula or its anisotropic generalisation \citep{cud91a}. This has the advantage that if the initial potential is close to the true (converged) potential -- in the sense of the iterative procedure described above -- then the density profile of each component in the converged model will be similar to the initial one; also, it will have the same total mass, but with a somewhat flattened shape. The downside of this type of DF is that it lacks a closed (analytic) form, but it can always be approximated to any desired accuracy by a smooth function that permits rapid computation (e.g. a cubic spline).

A Quasi-Isothermal DF -- appropriate for discy components -- is readily available in closed form \citep[see][his section 4.3, equation 20]{Vasiliev2019}, and a particular member of this family is defined by choosing appropriate parameter values. The parameters that are relevant to our modelling, as well as their values, are listed in Tab.~\ref{t:dfdisc}. We stress that any parameter that enters the definition of these DFs, but is not listed here, takes on its default value as defined by \agama\ (see Footnote \ref{fon:doc}).

Our main barred model is sampled at a very high resolution, identical to the resolution of our smooth model \citep[cf.][]{BlandHawthorn2021}: The DM halo is sampled with 20 million particles; the stellar bulge, with 4.5 million particles; and the stellar disc with 50 million particles. This implies a particle mass of roughly $5\times10^4$ \Msun\ (halo), $3\times10^3$ \Msun\ (bulge), and $9\times10^2$ \Msun\ (disc). We also create a `low-resolution' model for the purpose of comparison. In this model, the DM halo and the bulge are sampled with 10 times less particles compared to our main model, while the disc is sampled with 50 times less particles. As discussed in \citet[][]{BlandHawthorn2021}, we have established that our synthetic discs sampled with $N \ll 10^8$ particles are subject to {\em numerical} (i.e. artificial) heating due to the coarseness of the density interpolation, particularly in the outer regions. In contrast, we do not expect our disc sampled with $5\times10^7$ particles to  display such behaviour - at least within the time period consider here, and any apparent heating can be reasonably attributed to the bar formation process.\\

Fig.~\ref{f:ics1} (top) presents the observed
(data points) and model-predicted
rotation curve (gray solid line) of the Milky Way. 
The predicted circular velocity curve is computed from the azimuthal average of the potential.
The data points are taken from \citet[][masers]{rei19a}, \citet[][red giants]{eil19a} and \citet[][multiple sources]{Sofue2009}. The unevolded model (solid line) is in good agreement with Reid's and Eilers' data except that this is for the pure exponential disc model. Once the bar sets in (broken lines), it redistributes disc stars (black dashed line) such that the rotation curve peaks near the centre,
but drops and flattens near the solar circle, more in line with Sofue's data compilation.

This behaviour is consistent with \citet{Fujii2019}. In both our and their model, the rotation curves reaches a higher amplitude near the centre after the bar forms, relative to the initial rotation curve \citep[see also][]{Athanassoula1986,Rodionov2017}.

It is unclear however whether this is representative of the MW's rotation curve towards the centre \citet{Portail2017}. The data compilation by \citet[][]{Sofue2009} at first sight suggests that the Galactic rotation curve peaks at the centre. But this is not the case when the non-circular gas flow
in the potential of the bar is modelled hydrodynamically \citep[e.g.,][]{Chemin_2015,Li_2021}.


Fig.~\ref{f:ics1} (middle and bottom) reveal important aspects of how our model is set up to produce a bar. The smooth, non-evolving HBD model has a disc-to-total mass fraction $f_{\rm d}$, at a given radius, that never exceeds the threshold value (middle panel, horizontal dotted line). But $f_{\rm d}$ for the smooth model {\it is entirely unrepresentative of what we know to be true in the Milky Way}, as shown by the cross. Once we raise $f_{\rm d}$ to its correct value by the appropriate choice of disc mass (solid orange curve), {\em a bar is formed with the right properties}, as we discuss later. We consider this to be a remarkable outcome arising from these new simulations. Note that the value of $f_d$ in our model remains consistent with the Milky Way value {\em after} the bar forms (broken curves).

\subsubsection{The impact of Toomre's $Q$ criterion}

\begin{figure}
\centering
\includegraphics[width=0.45\textwidth]{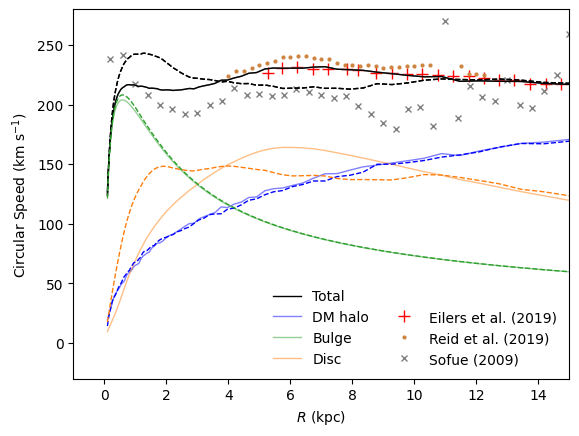}
\includegraphics[width=0.45\textwidth]{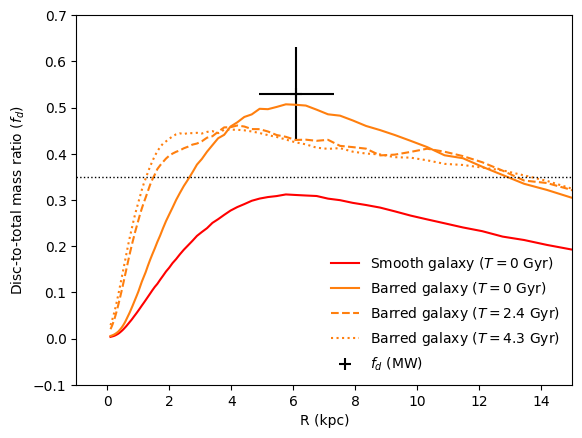}
\includegraphics[width=0.45\textwidth]{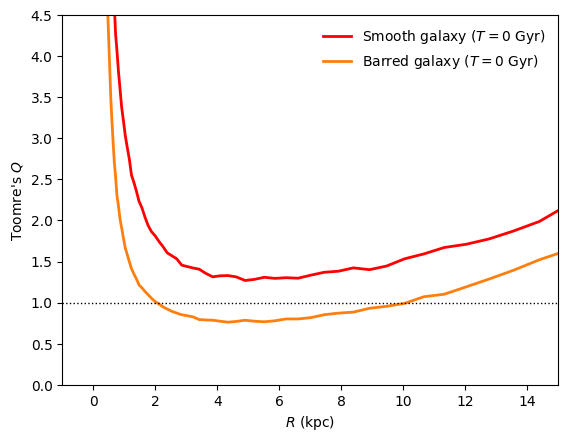}
\vspace{-5pt}
\caption[  ]{Top: The observed circular velocity compared to our model's result. The data points are from \citet[][brown `$\cdot$' symbols]{rei19a},  \citet[][red `+' symbols]{eil19a}, and \citet[][gray `$\times$' symbols]{Sofue2009}. The latter have been binned over $R = [0, 20]$ kpc using 50 bins of uniform size.  The model is shown at $T = 0$ Gyr (solid curves) and after the bar is fully formed at $T \approx 2.4$ Gyr (dashed curves). The green, orange, and blue curves correspond, respectively, to the bulge's, the disc's, and the halo's contributions. The 
redistribution of matter before and after the bar is formed is reflected in the disc component, and only minimally in the bulge and dark halo.
Middle: Disc-to-total mass ratio.  Barred galaxy model at $T = 0$ Gyr (solid curve), at $T \approx 2.4$ Gyr (dashed curve), and at $T \approx 4.3$ Gyr (dotted  curve). The horizontal, dotted black line indicates the threshold $f_{\rm d} \sim 0.35$ (see text for details). The datum (+) indicates the range of estimates for the MW \citep{blandhawthorn2016}. Bottom: Toomre's $Q$ parameter. The corresponding value for a smooth galaxy at $T = 0$ Gyr (solid red curve) is included for comparison.
}
\label{f:ics1}
\end{figure}

The Toomre $Q$ criterion measures stability against {\it local axisymmetric} perturbations in the presence of shear from differential rotation and internal random motions, such that 
\begin{equation}
    Q = \frac{\kappa ~\sigma_R}{a ~G \Sigma} \, ,
\end{equation}
where $\kappa$ is the in-plane epicyclic frequency, $\sigma$ is the internal dispersion, $\Sigma$ is the local surface density of the disc. The constant factor $a=\pi$ for a fluid analysis, and $a=3.36$ for our distribution function approach \citep{too64a}.
The stellar disc in our model is characterised by $Q \lesssim 1$ within $2.5 \lesssim R / \kpc \lesssim 10$ (Fig. \ref{f:ics1}, bottom); this is a deliberate choice to speed up the onset of the bar instability \citep{Fujii2018}. Thus the $Q$ criterion is not directly implicated in bar formation, but the bar onset timescale does have a secondary dependence on $Q$.

%

\subsubsection{Matching the Galactic parameters} \label{s:match}

\begin{figure}
\centering
\includegraphics[width=0.45\textwidth]{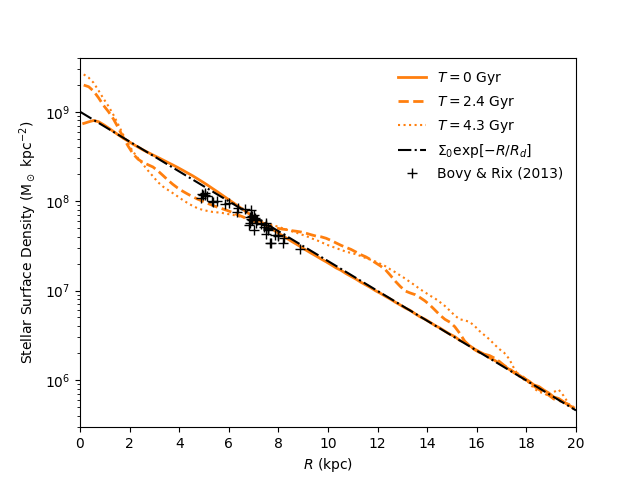}
\includegraphics[width=0.45\textwidth]{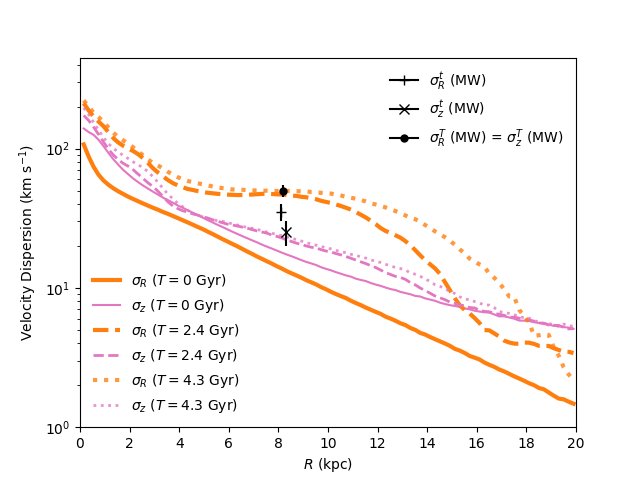}
\vspace{-5pt}
\caption[  ]{Top: Azimuthally averaged surface density profile of the stellar disc at $T = 0$ Gyr (solid  curve), at $T \approx 2.4$ Gyr (dashed curve), and at $T \approx 4.3$ Gyr (dotted curve). The dot-dashed black curve indicates a classical exponentially declining profile,  $\Sigma_0 \exp \left[ -R / R_d \right]$ with $\Sigma_0 = 10^9 \Msun \kpc^{-3}$ and $R_d  = 2.6$ kpc. Data ($\boldsymbol{+}$) from \citet{bov13a}, corrected for the contribution of gas and dark matter to the local acceleration (see text). Bottom: Azimuthally averaged profile of the radial velocity dispersion (thick lines) and vertical velocity dispersion (thin lines) of the stellar disc at $T = 0$ Gyr (solid curves), at $T \approx 2.4$ Gyr (dashed curves), and at $T \approx 4.3$ Gyr (dotted curves). For reference, we include the corresponding values for the Milky Way at the solar circle ($R_0 = 8.2$ kpc) for the {\em thin} disc: $\sigma^t_R = 35 \pm 5$ \kms\ ($\boldsymbol{+}$); $\sigma^t_z = 25 \pm 5$ \kms\ ($\boldsymbol{\times}$); and for the {\em thick} disc: $\sigma^T_R = \sigma^T_z = 50 \pm 5$ \kms\ ($\bullet$). The data points have been shifted by $\Delta R = \pm 0.1$ kpc around $R_0$ for clarity. Velocity dispersion data from \citet{blandhawthorn2016}. }
\label{f:ics2}
\end{figure}

Our choice of parameter values defining the density profile of the dark halo, the bulge, and the disc, in addition to the parameter values defining the disc's DF, yields a model where:
\begin{itemize}
    \item The halo and the disc both have a mass and a structure consistent with the corresponding properties for the MW; this is also true for the barred (i.e. evolved) model.
    \item The stellar disc follows
    an exponentially declining, surface density profile with an amplitude and scalelength consistent with the corresponding estimates for the MW (Fig.~\ref{f:ics2}, top); it is well described by $\Sigma_0 \exp{\left[ -R / R_d \right]}$ with $\Sigma_0 = 10^9$ \Msun\;  \kpc$^{-2}$ and $R_d = 2.6$ kpc.
    \item The (azimuthally averaged) radial and vertical velocity dispersions of the disc at the solar circle are in good agreement with the MW's values (Fig.~\ref{f:ics2}, bottom).
    \item The model-predicted circular-velocity curve of the barred model as shown in Fig.~\ref{f:ics1} (top) seems reasonable overall, although it appears too high within the bar region.
    \item The disc-to-total mass ratio (Eq. ~\ref{e:fd}) is well above the threshold for the onset of the bar instability within a Hubble time, $f_{\rm d} \approx 0.35$, and it agrees well with the MW estimate within the uncertainties (Fig.~\ref{f:ics1}, middle).
\end{itemize}

We note that the surface density of the synthetic stellar disc (Fig.~\ref{f:ics2}, top) compares well with the measurements of the local vertical acceleration by \citet{bov13a}. Since the stellar disc alone induces a smaller acceleration, which results in a lower inferred surface density, we have corrected their measurements assuming an equal contribution from baryons and dark matter, in line with the MW's $f_d \approx 0.5$.


%

\begin{table}
\begin{center}
\caption{ Properties of the barred galaxy model and comparison to the MW. The values of the model are given at $T \approx 2.4$ Gyr. With exception of $\Omega_p$,
all quantities maintain the values given here throughout the evolution of the model out to $T \approx 4.3$ Gyr.
}
\label{t:comp}
\begin{tabular}{llcc}
Component & Observable & Model & MW \\
\hline
DM halo         & & & \\
                & $M_{vir}$ ($10^{12}$ \Msun)   & 1.18 & $1.3 \pm 0.3$ \\
                & $R_{vir}$ (kpc)               & 250 & $282 \pm 30$ \\
Stellar disc    & & & \\
                & $M_{tot}$ ($10^{10}$ \Msun)   & 4.3 & $3.7 \pm 1$$^b$ \\
                & $R_d$ (kpc)                   & 2.6 & $2.6 \pm 0.5$ \\
                & $h$ (kpc)                     & 0.3 & $0.3 \pm 0.05$$^a$ \\
                & $\sigma_R$ (\kms)$^{a,b}$     & 46 & $35 \pm 5$$^c$ \\
                & &                             & $50 \pm 5$ \\
                & $\sigma_z$ (\kms)$^{a,b}$     & 23 & $25 \pm 5$$^c$ \\
                & &                             & $50 \pm 5$ \\
                & $\Sigma_*$ (\Msun \pc$^{-2}$)$^a$ & 38.4 & $38 \pm 4$ \\
                & $f_{\rm d}$$^{,g}$     & 0.45 & $0.53 \pm 0.1$ \\
                Bar & & & \\
                 & Size (kpc)              & 4.5   & $5.0 \pm 0.2$ \\
                 & $\Omega_p$ (\kms~\pkpc) & & \\
                 & \hspace{20pt}at 2.4 Gyr     & 43      & 38$^e$ \\
                 & \hspace{20pt}at 4.3 Gyr     & 39       & 38$^e$ \\
                 & $\dot{\Omega}_p$ (\kms~\pkpc~\pGyr) & -2.3       & $-4.5 \pm 1.5$$^d$ \\
                Bar region &&&\\
                $R \lesssim 5$ kpc &&\\
                & Stellar mass$^f$ ($10^{10}$ \Msun)  & 3.5         & 3.17$^h$ \\
                $R \lesssim 2$ kpc &&&\\
                & Stellar mass ($10^{10}$ \Msun)  & 2.3   & -- \\
                & Dynamical mass ($10^{10}$ \Msun)  & 2.6   & 1.85$\pm$0.05$^h$ \\
                & Baryon fraction                   & 0.88  & 0.83$\pm$0.15$^h$ \\
\end{tabular}
\end{center}
\begin{list}{}{}
\item Notes: $^{a}$Measured at $R_\odot = 8.2$ kpc.
$^{b}$We do not distinguish between a thin and a thick disc component in our model; the MW disc mass quoted here includes an additional contribution of $6 \pm 3 \times 10^9$ \Msun\ from the thick disc. $^c$ Values corresponds to the MW's thin disc; the values below, to the thick disc. $^d$ \citet{Chiba2021}. $^e$Average value over a number of observations (see Fig.~\ref{f:a2o}, bottom right). $^f$Includes the long bar, the classical bulge, and the inner disc.
$^g$Measured at $2.2 R_d \approx 5.7$ kpc. $^h$ \citet{Portail2017}. 
All values for the MW are adopted from \citet{blandhawthorn2016} unless indicated otherwise.
\end{list}
\end{table}

\begin{figure*}
\centering
\includegraphics[width=\textwidth]{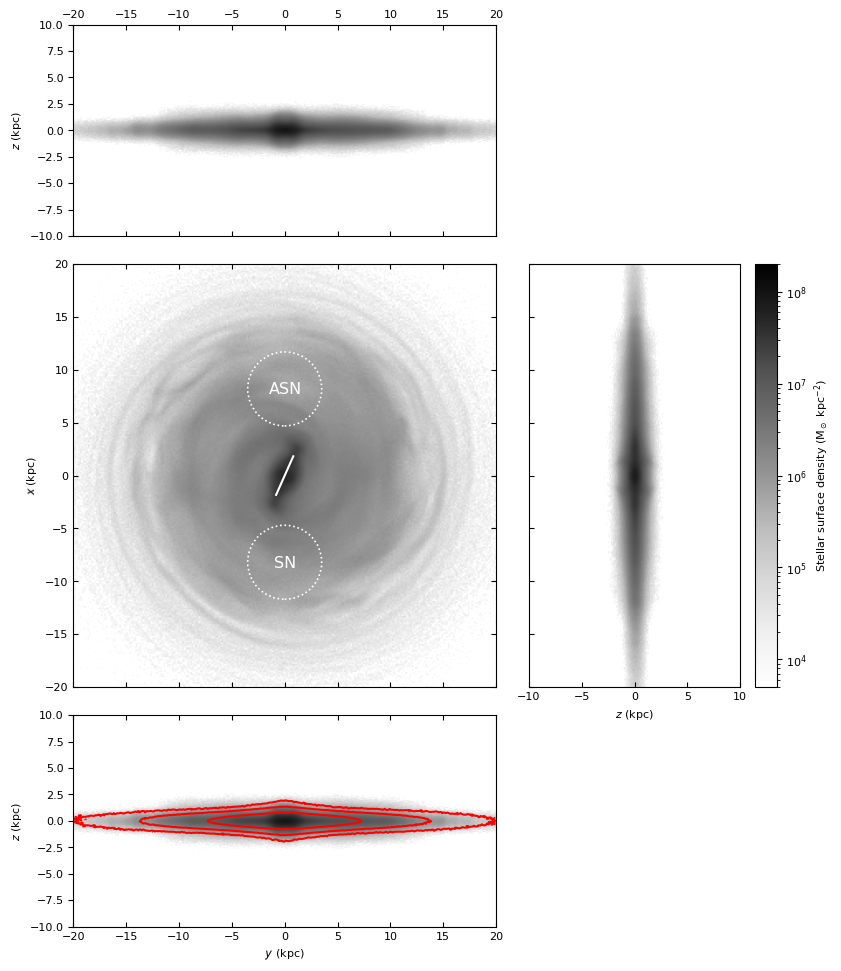}
\vspace{-5pt}
\caption[  ]{ Surface density of the stellar disc at $\sim 2.4$ Gyr when the bar is fully formed. Note that the classical bulge is {\em not} included in either projection. Top: $y - z$ projection. Middle left: $y - x$ projection. The approximate inclination of the bar with respect to the {\em negative} $x$-axis indicated by the white solid line is $\sim$24\deg. The white, dotted circles indicate the edge of the spherical volumes enclosing the stars used sample the kinematic properties of stars (see Figs.~\ref{f:vphi_R}, \ref{f:kin}) in the Solar neighbourhood (SN) and the Antipodal SN (ASN; see text for details). 
Middle right: $z - x$ projection. The box/peanut shape of the bar in this projection is apparent (cf. Fig.~\ref{f:surfdens_bar2}). The bottom sub-panel is identical to the top sub-panel, but includes a series of contours (red) corresponding to a surface density of $10^5$, $10^6$, and $10^7$ \Msun $\kpc^{-2}$ which show the initial state of the disc, for comparison. Clearly, although the overall mass distribution of the disc changes only modestly from its initial state to the bar formation epoch, a thickening of the disc within $R \approx 15$ kpc at the lowest densities shown is visible. An animation of the full evolution is available (see Sec.~\ref{s:data}). 
}
\label{f:surfdens_bar}
\end{figure*}

\section{Simulations} \label{s:sim}

\subsection{\ramses\ setup}\label{s:setup}

We have discussed our typical set-up at length in earlier papers \citep{Tepper-Garcia2018,BlandHawthorn2021}. Here we provide only a brief outline. 
The compound system (DM halo - bulge - disc) is evolved with the \ramses\ code \citep{tey02a}, which incorporates adaptive mesh refinement (AMR). The system is placed into a cubic box spanning 600 kpc on a side. The AMR grid is maximally refined up to level 14, implying a limiting spatial resolution of \mbox{$\delta x$ = 600 kpc / $2^{14} \approx 36$ pc}. The maximum resolution is achieved everywhere within $R \leq 10$ kpc, the most relevant spatial domain of our study.
Due to the presence of the simulation mesh, the force exerted onto any particle is necessarily softened. In \ramses, the force softening is on the order of a length equivalent to the side of cubic volume element (cell) $\delta x$. In our simulation, this scale is roughly an order of magnitude smaller than the typical disc scaleheight, and well below the typical size of the emergent bar (see Sec.~\ref{s:heat}).

The total simulation time is roughly 4.3 Gyr. As explained below, the bar is fully formed after about 2.4 Gyr, implying a total simulation time of the barred galaxy of $\sim 2$ Gyr, a time span long enough to allow the investigation of the bar's long-term evolution.

\begin{figure*}
\centering
\includegraphics[width=\textwidth]{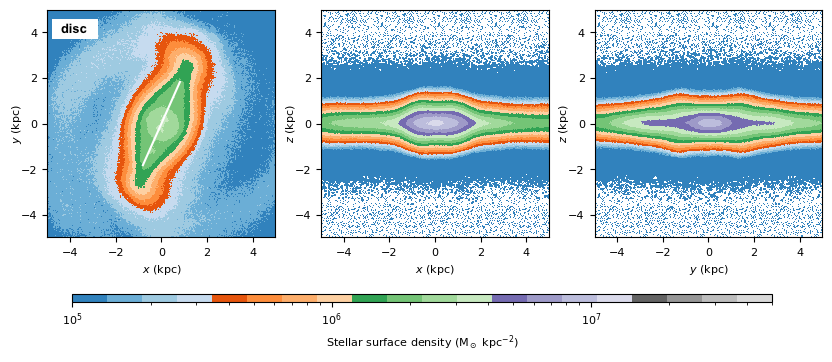}\\
\includegraphics[width=\textwidth]{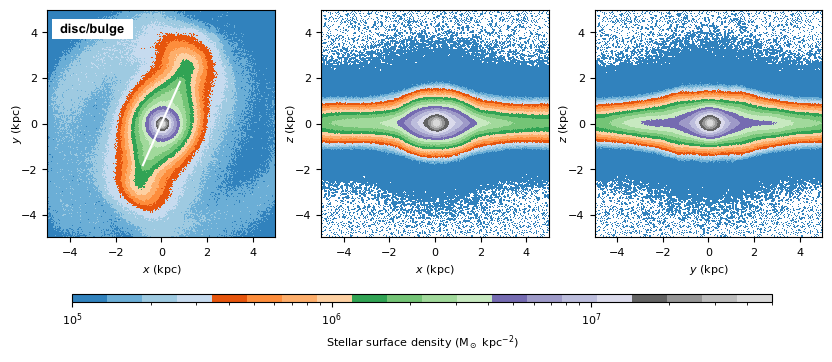}
\caption[  ]{ Top. Surface density of the synthetic stella disc within a cubic volume of 5 kpc per side centred on the bar, along three orthogonal projections: $x - y$ (left), $x - z$ (centre), and $y - z$ (right). Compare to Fig.~\ref{f:surfdens_bar}, which displays the same projections but across a larger spatial scale. Note that densities below $10^5$ \Msun\ kpc$^{-2}$ are all encoded using the same colour (blue). It is worth emphasising that the thick component apparent in the central and right panels at $|z| > 1$ kpc is extremely low density (cf. Fig.~\ref{f:surfdens_bar}, top and bottom sub-panels). Bottom. Same as the top row, but including the classical bulge.  This figure shows an orthogonal projection, in contrast to Fig. \ref{f:surfdens_bar3} which displays a side elevation view of the bar region.}
\label{f:surfdens_bar2}
\end{figure*}

\begin{figure}
\textcolor{white}{~}\hfill 
\includegraphics[width=0.45\textwidth]{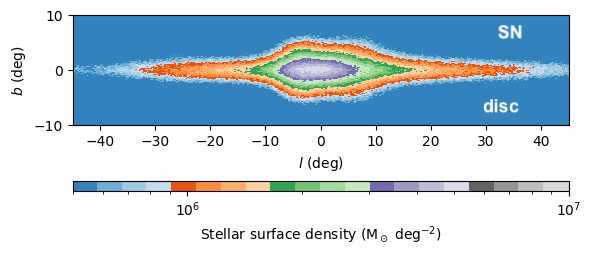}\\
\textcolor{white}{~}\hfill 
\includegraphics[width=0.45\textwidth]{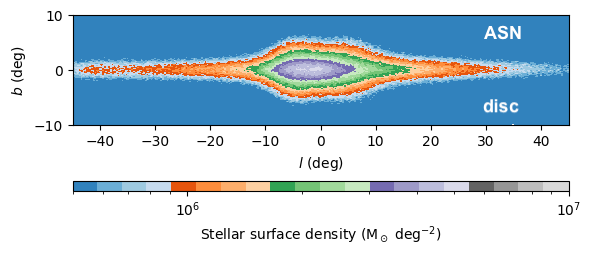}\\
\textcolor{white}{~}\hfill 
\includegraphics[width=0.45\textwidth]{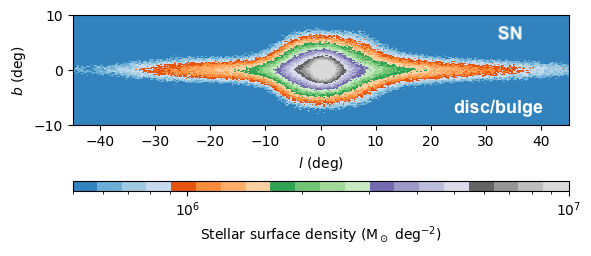}\\
\textcolor{white}{~}\hfill 
\includegraphics[width=0.45\textwidth]{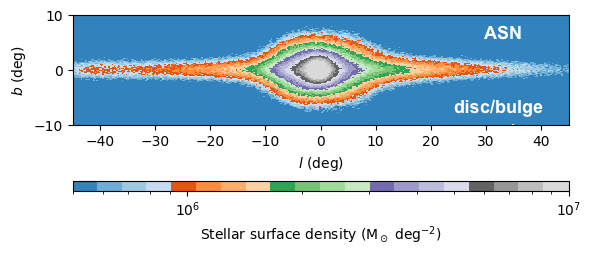}\\
\textcolor{white}{~}\hfill 
\includegraphics[width=0.43\textwidth]{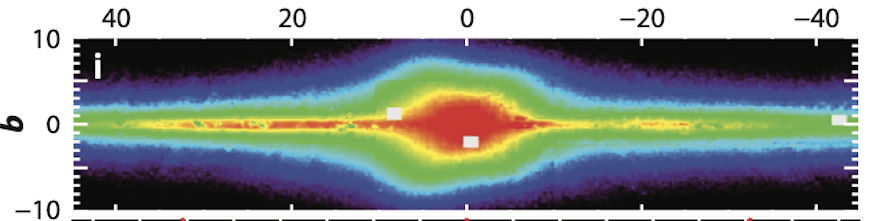}
\caption[  ]{ Heliocentric view (side elevation) of the bar region. The top four panels correspond to our synthetic galaxy. We assume a Sun-Galactic Centre distance of 8.2 kpc. The top two panels include only the stellar disc; the next two panels include the disc and the classical bulge. `SN' indicates the view from the Solar neighbourhood while `ASN', the view from the SN antipode (see Fig.~\ref{f:surfdens_bar} for a definition of the SN and ASN volumes). Bottom panel: Milky Way model from \citet{Wegg2015}. Compare to Fig.~\ref{f:surfdens_bar2}, central panel, which provides an orthogonal projection. }
\label{f:surfdens_bar3}
\end{figure}

\section{Results} \label{s:res}

In what follows, we focus on our high-resolution barred model, with only occasional references to the smooth model or the low-resolution barred model. 
This section demonstrates that our N-body model is an appropriate surrogate for the Milky Way in terms of its internal structure and dynamics.

To illustrate the quality of our MW surrogate, we present in Fig.~\ref{f:surfdens_bar} three projections of the stellar disc at $T = 2.4$ Gyr when the bar reaches its peak strength (see Sec.~\ref{s:bar2}).
Since the bar maintains its strength for at least 2 Gyr, we get similar results at later times and therefore this snapshot is representative.

To further emphasise the rich structure of the central region, in particular of the bar, we present Fig.~\ref{f:surfdens_bar2} (top), essentially a zoomed-in version of Fig.~\ref{f:surfdens_bar}. In the same figure (bottom) we include a series of analogous projections but including the classical bulge. Finally, to aid the comparison to observations, we provide an heliocentric view of the bar in Fig.~\ref{f:surfdens_bar3}, both for the synthetic galaxy and the Milky Way, both with and without the classical bulge.

It is worth noting that, given the bisymmetry imposed by the bar, there are two equivalent volumes that we consider to be representative of the Solar neighbourhood (SN). In our model, these are located at $x = -8.2$ kpc and $x = 8.2$ kpc (see Fig.~\ref{f:surfdens_bar}). We refer to these as SN and the antipodal Solar neighbourhood (ASN). For the SN region, the simulation has $\sim 1.7\times10^6$
particles, roughly a quarter of the of the number of stars in the \gaia\ RVS over the same volume. This compares with  $\sim 2.2\times10^6$
particles in the antipodal volume, or roughly one third of the Gaia sample.

As a final illustrative example, we provide a collage in the Appendix (Figs. ~\ref{f:collage} and \ref{f:collage2}) comparing our model with the Galaxy and NGC4314, which \citet{blandhawthorn2016} selected because it shows a similar face-on morphology as the MW morphology inferred from star counts and modelling.

A quantitative assessment of our model is provided in Tab.~\ref{t:comp}. There we contrast the estimated values for an array of MW observables with the corresponding results retrieved from our evolved, barred MW surrogate. The latter are presented at $T \approx 2.4$ Gyr. Given the long-term stability of the N-body model, these values are representative of the model out to $T \approx 4.3$ Gyr, i.e. over a time span of $\sim$ 2 Gyr. The notable exception is the bar pattern speed, which decreases with time, in agreement with recent estimates of the bar slow down \citep{Chiba2021}. Overall, the values from the model agree remarkably well with the corresponding MW values.

\subsection{Stellar kinematics} \label{s:rc}

\subsubsection{Global signatures}

\begin{figure*}
\centering
\includegraphics[width=0.45\textwidth]{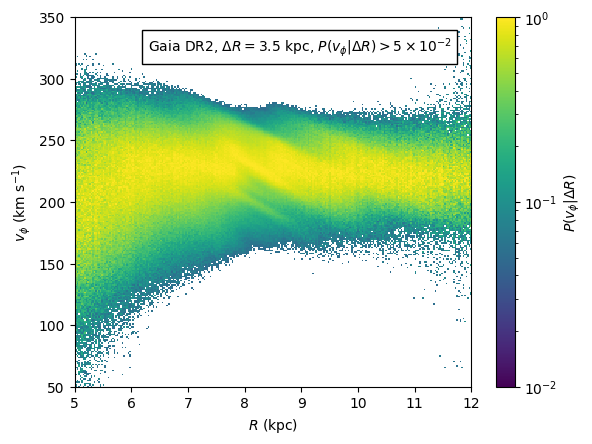}
\includegraphics[width=0.45\textwidth]{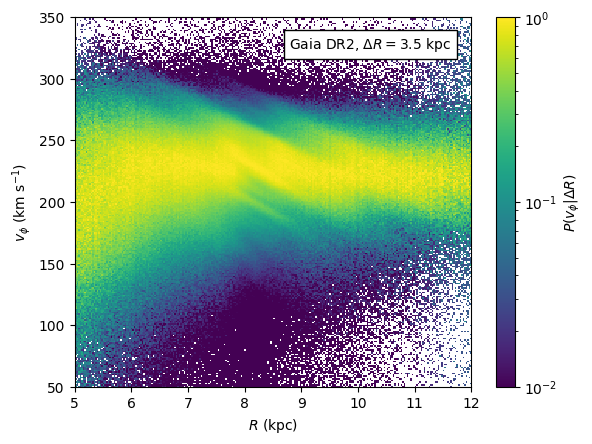}
\includegraphics[width=0.45\textwidth]{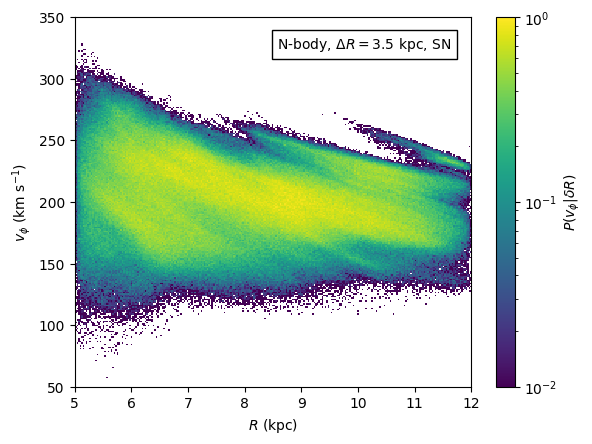}
\includegraphics[width=0.45\textwidth]{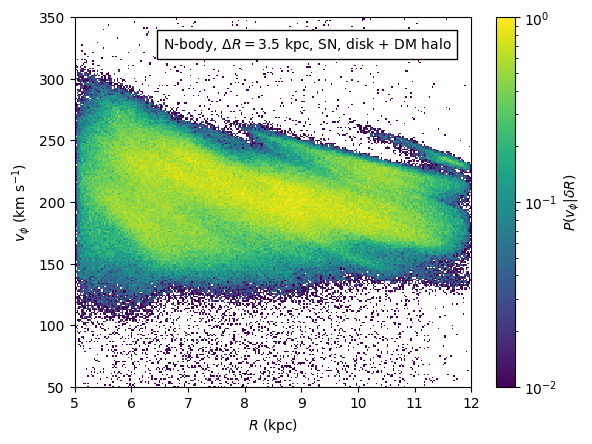}
\includegraphics[width=0.45\textwidth]{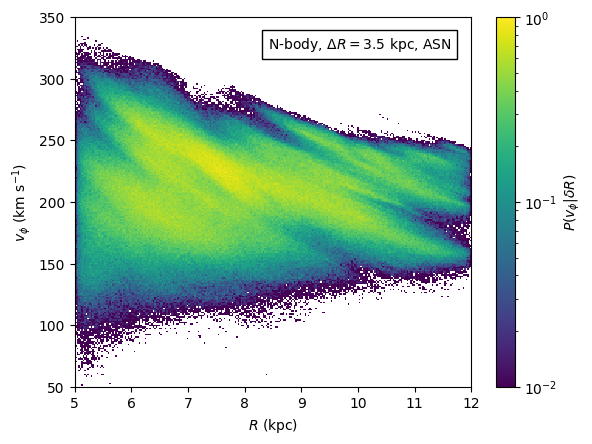}
\includegraphics[width=0.45\textwidth]{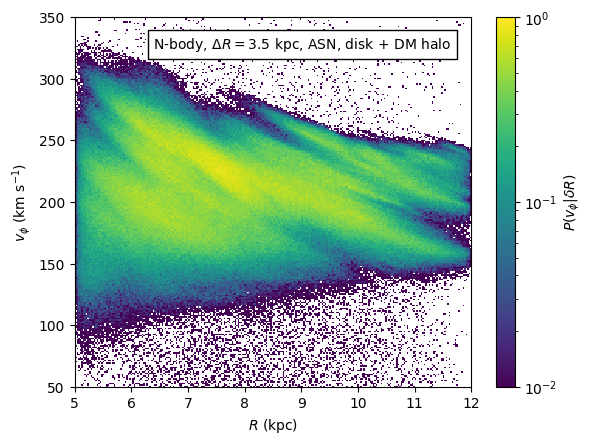}
\vspace{-5pt}
\caption[  ]{Top left: Distribution of \gaia\ DR2 stars on the $(v_\phi, R)$ plane selected in a sphere of radius 3.5 kpc centred at $R_\odot = 8.2$ kpc. The distribution is represented as a conditional probability, $P(v_\phi | \delta R)$. The peak of the distribution has been normalised to unity for each radial bin $\delta R$ to remove the selection effect inherent to $R$. 
We also impose a $(v_\phi, R)$ density cut-off, $P(v_\phi | \delta R) > 5\times 10^{-2}$, to aid comparison with the model (middle, bottom). The density threshold removes 3\% of stars in the RVS.
Top right: The same as the top left figure without the density threshold showing all RVS stars.
Middle left: Distribution of stars on the $(v_\phi, R)$ plane in the synthetic stellar disc selected in the same way in the SN volume (indicated by the bottom white circle in the central panel of Fig.~\ref{f:surfdens_bar}). The peak of the distribution has been normalised in the same way as the RVS data. The outer envelope corresponds reasonably well to the \gaia\ RVS data for the majority of stars.
Middle right: The same as the middle left figure but with dark halo particles included to illustrate the impact of a stellar halo (spheroidal) component in the $(v_\phi, R)$ plane. The missing 3\% of stars are not well represented by our {\it isolated} barred disc model.
Bottom panels: Similar to the corresponding panels in the middle row, but for the antipodal SN volume (ASN; indicated by the top white circle in the central panel of Fig.~\ref{f:surfdens_bar}).
}
\label{f:vphi_R}
\end{figure*}

While the circular rotation curves in Fig.~\ref{f:ics1} are the traditional approach to comparing models and data, we can go further in an era of ESA \gaia. The Radial Velocity Survey (RVS) allows us to compute the full 6D phase space of about six million stars within roughly 3.5 kpc of the Sun \citep{Antoja2018}. We illustrate the predictive power of simulated Galaxy models by comparing the distribution in the $R$ vs. $V_\phi$ plane as seen by the RVS (Fig.~\ref{f:vphi_R}, top) with our simulated version of the same survey: SN (middle) and ASN (bottom). 
The extracted regions corresponding to the latter are shown as white circles with diameter 7 kpc in the $y-x$ plane (Fig.~\ref{f:surfdens_bar}, main panel).
In our synthetic reconstruction, a simple selection function is used based on a star's distance from the Sun, as discussed in the caption.

In Fig.~\ref{f:vphi_R} (top right), the outer envelope of the cells defined by the conditional probability $P(V_\phi\vert \delta R) > 0.05$ is comparable across all panels; outside of this region, there are only a few stars per cell. To underscore this point, we plot the $R-V_\phi$ diagram for stars in RVS with $P(V_\phi\vert \delta R) > 0.05$ (top left), which selects out the broader envelope. 

We note that all stars from the original RVS survey are included with no selection on age, kinematics or chemistry. In this plane, a ``rotation curve'' is not apparent because most stars in the Galaxy undergo precessing, 3D non-circular orbits within a few rotation periods after being born. There is diagonal banding in both the RVS data and our model due to spiral arms and resonances \citep[e.g.][]{Khanna2018},  but we stress that we do not expect these to align with any degree of precision without the presence of a self-gravitating gas phase. This is addressed in our follow-up paper. None the less, the overall agreement between our model and the MW is encouraging.


%


\begin{figure*}
\centering
\includegraphics[width=\textwidth]{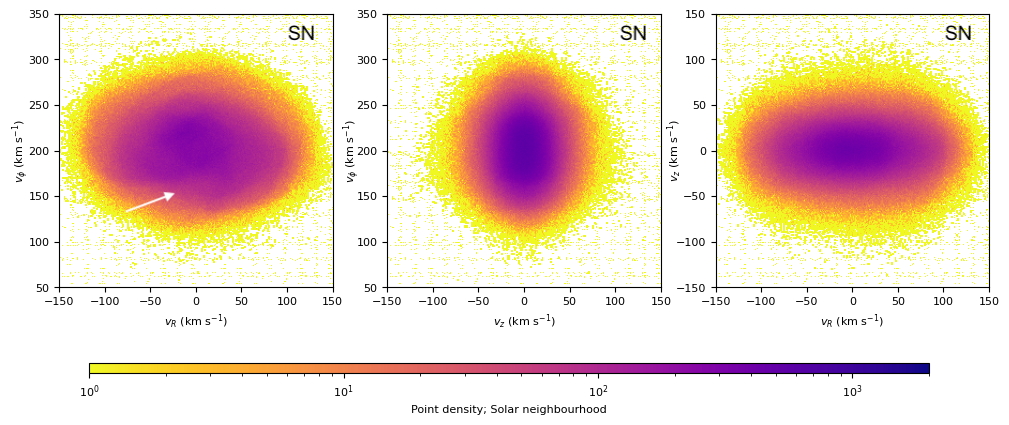}\\
\includegraphics[width=\textwidth]{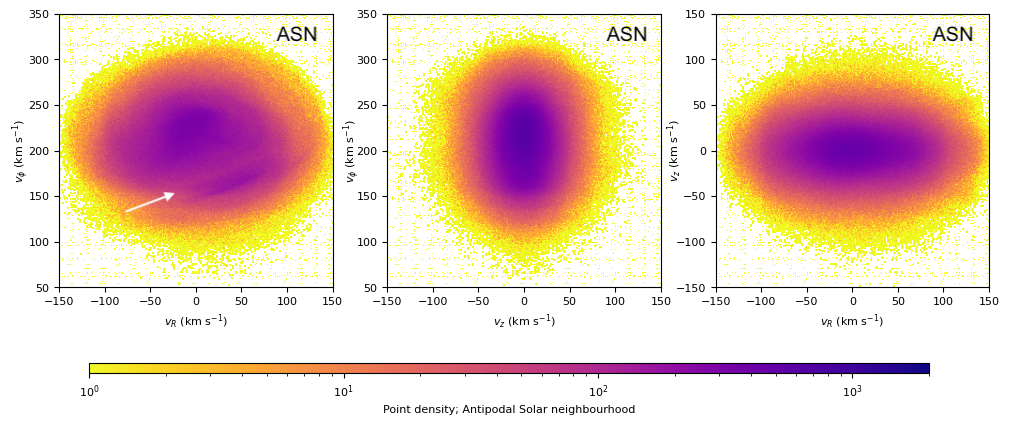}
\vspace{-10pt}
\caption[  ]{Stellar kinematics around the Sun in different phase planes. Left: $(v_R,v_\phi)$. Middle:  $(v_z,v_\phi)$. Right: $(v_R,v_z$). The top row corresponds to the Solar neighbourhood (SN); the bottom row, to the Antipodal Solar neighbourhood (ASN). For the definition of `SN' and `ASN', see Fig.\ref{f:surfdens_bar}. The white arrow in each of the right panels points to the diagonal discontinuity apparent in the distribution (see Sec.\ref{s:kin} for further details).
}
\label{f:kin}
\end{figure*}

\begin{figure*}
\centering
\includegraphics[width=\textwidth]{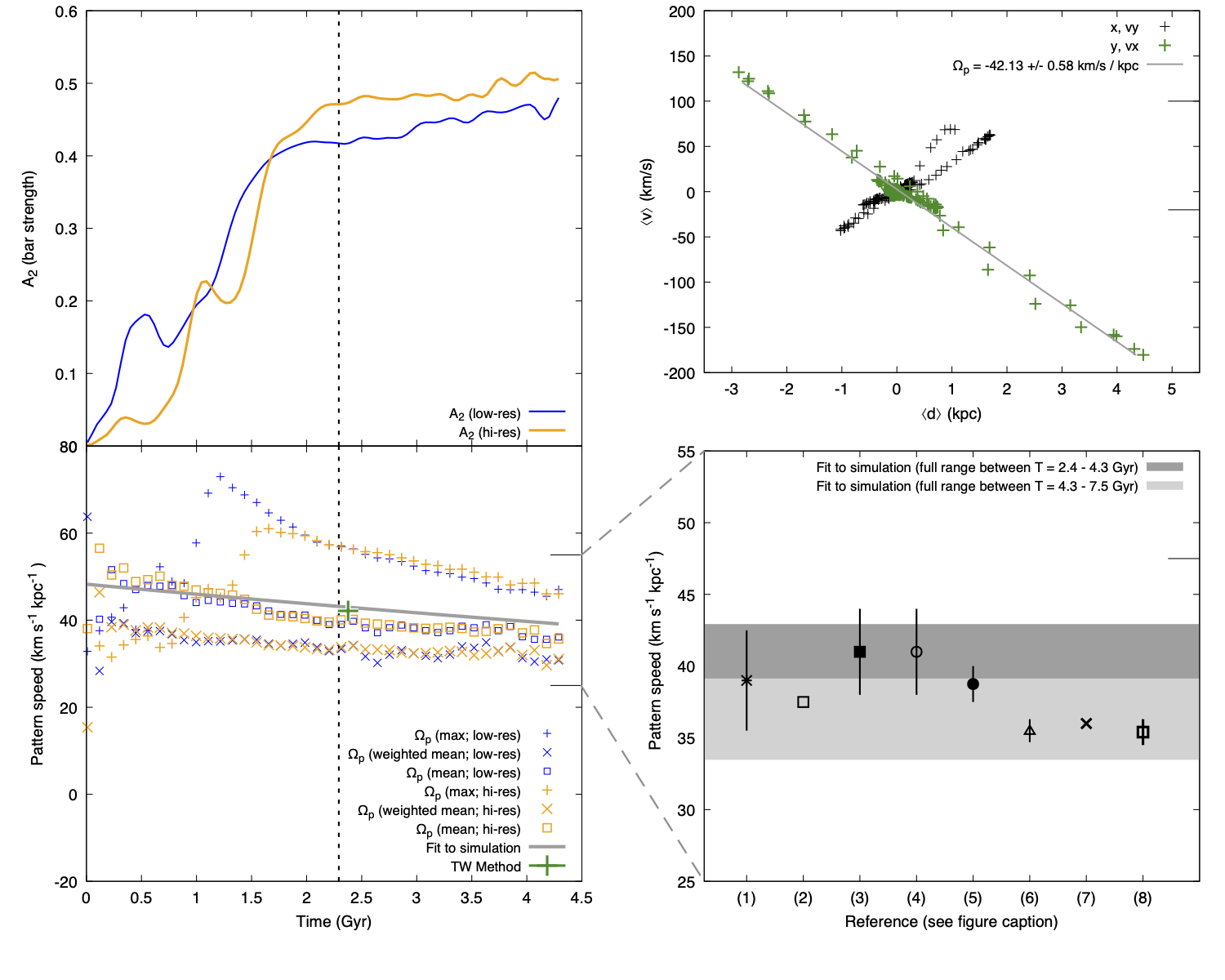}
\vspace{-15pt}
\caption[  ]{ Top left: Evolution of the bar strength with cosmic time. The bar strength is quantified by the maximum amplitude of the $m=2$ mode, $A_2$, within the bar region. Note how $A_2$ rapidly climbs up to $\approx 0.4 - 0.5$ by $T \approx 2.4$ Gyr, and then flattens out, indicating that the bar is fully formed and stable at this epoch. Thus we take the bar formation epoch to be at $\sim 2.4$ Gyr, flagged by the vertical dotted line (slightly shifted to the left for clarity). Top right: Bar pattern speed from \citet{tre84h}'s method in a selected snapshot of our simulation. The green symbols indicate the measurement perpendicular to the bar major axis; the black symbols, to the measurements parallel to the bar major axis. Note that $\langle d \rangle = 0$ kpc corresponds to the bar centre. Bottom left: Bar pattern speed from Fourier analysis. Comparison of different methods to estimate the pattern speed. We include the datum (green '+' symbol) obtained from the TW method (shown on the top right). The fit to the bar pattern speed in the simulation is given by an exponentially declining function $\Omega_p(T) = \Omega_p(0) \exp \left[ -T / \tau \right]$. See text for details. Bottom right: Pattern speed measurements for the MW bar. The dark grey band corresponds to the fitted $\Omega_p$ curve shown in the bottom left panel over $T \approx 2.4 - 4.3$ Gyr; the light grey band corresponds to the extrapolation of the fit out to $T=7.5$ Gyr. Key to references: (1) \citet{Portail2017}; (2) \citet[][no error bar]{cla19k}; (3) \citet{bov19n}; (4) \citet{san19v}; (5) \citet{li21e}; (6) \citet{Chiba2021}; (7) \citet[][no error bar]{Binney2020}; (8) \citet{cla21m}. }
\label{f:a2o}
\end{figure*}

\subsubsection{Stellar kinematics in the Solar Neighbourhood} \label{s:kin}

In Fig.~\ref{f:kin}, we present three different phase planes for stellar particles in both the SN (top) and ASN (bottom) volumes. In each case, the extent of the elliptic envelope (FWHM) compares well with the locally observed distribution \citep{Dehnen1998}, particularly as seen by the \gaia\ RVS \citep{Antoja2018}.

In the $(v_R,v_\phi$) plane, the diagonal discontinuity and ridgeline indicated by the arrow at $v_\phi \approx 150$ \kms\ is similar to the same structure observed in the Milky Way \citep{Gardner2010,Antoja2014,Quillen2018}. In particular, the linear ridgeline is reminiscent of the Hercules Stream that is believed to be associated with the bar corotation resonance 
\citep{Perez-Villegas2017,Binney2020}. 
The break is more evident in the ASN volume where there are 27 percent more disc particles compared to the diametric volume.\\

We have addressed aspects of the MW that are well explained by our model. But there are new observations that we would not expect to reproduce in an isolated disc model. In particular, we do {\it not} detect any evidence of substructure, specifically the phase spiral \citep{Antoja2018}, in the $(z,v_z)$ plane. Numerous studies have now shown that this beautiful phenomenon is most likely associated
with the interaction of the Sgr dwarf with the Galactic disc \citep[see][for a recent review]{Brown2021}. There is
no such interaction in the isolated disc model presented here, or in \citet{Fujii2019}.
But both models are able to recover the phase-spiral signature when the disc is perturbed by a
transiting mass \citep{Gandhi2021,BlandHawthorn2021}.

\subsubsection{Disc heating} \label{s:heat}

In Fig.~\ref{f:ics2} (bottom), we have plotted the vertical and radial dispersion in log-scale to allow for easy comparison with \cite{Fujii2019}. Their largest simulation has $8\times$ more disc particles and runs for at least twice as long compared to ours, but the results from both teams compare remarkably well, both qualitatively and quantitatively. We both observe the radial dispersion $\sigma_R$ to grow dramatically and to extend further in radius with time; this tracks the growth of the bar. In comparison,
we see low-level vertical heating of the disc, as measured by $\sigma_z$, also in agreement with \cite{Fujii2019}. The heating effect grows with radius, again as a consequence of the bar growth. It is remarkable that there is so little leakage from radial to vertical modes, strongly suggesting that the bar action is mostly confined to the plane.
This contrasts with claims of strong vertical heating or buckling occasionally observed in earlier N-body simulations \citep[e.g.][]{Saha2014}, although we note that these are mostly low resolution ($N\lesssim 10^7$ particles).
Although we have already demonstrated that numerical heating is a problem in this limit using a non-barred galaxy \citep{BlandHawthorn2021}, it is unclear to what extent a low-particle number contributes to the vertical heating in a synthetic galaxy that eventually develops a bar.
 
As mentioned above, we see a factor of 2$-$2.5 increase in $\sigma_R$ when averaged over the solar annulus after the bar has formed (Fig.~\ref{f:ics2}), as do \citet{Fujii2019}. This is a potential discrepancy with observations of the Milky Way, except to note that the enhanced radial dispersion matches the observed value for the $\alpha$-rich disc \citep[historically, the `thick disc';][]{blandhawthorn2016}. 
Some of the difference may reflect the way in which reliable measurements of $\sigma_R$ in the Milky Way are restricted to local volumes in the solar neighbourhood \citep{sharma2014,Piffl2014}.\\

We close this section by stressing that the results presented in Figs.~\ref{f:ics1} (top), \ref{f:ics2}, \ref{f:surfdens_bar}, and \ref{f:kin} agree extremely well with the results corresponding to the best model by \citet[][dubbed `MWa5b']{Fujii2019}. This circumstance is far from obvious and it is quite remarkable as it is encouraging, given the very different start-up codes, the different N-body solvers, the different computer architectures, and the vastly different resolutions (i.e. particle numbers) adopted by each team. We believe that this is testament to how well both GalactICs and \agama\ are able to set up an self-consistent configuration at the outset, as it is to the consistency of the system's evolution computed with their tree-based code and the grid-based \ramses.

\subsection{Bar evolution} \label{s:bar}

\subsubsection{Bar instability onset and growth} \label{s:bar2}

The contents of the middle panel in Fig.~\ref{f:ics1} lie at the heart of our new research. The red curve shows the dependence of $f_{\rm d}$ with radius for the smooth disc model. This ratio never exceeds the threshold (shown as a horizontal dotted line) identified by \citet{Fujii2018}, and therefore a bar cannot emerge within a Hubble time (Sec.~\ref{s:intro}). But the Galactic value of $f_{\rm d}$ is almost a factor of two higher (indicated by the cross). Once the value of $f_{\rm d}$ is increased in our model to be consistent with the Galactic value, indicated by the solid orange curve, the disc becomes bar unstable in a relatively short time and, as we show next, {\it produces a bar that is remarkably similar to what we observe.}

To illustrate the impact of $f_{\rm d}$, we provide movie sequences for both the smooth and barred models (see Sec.~\ref{s:data}). In the barred model (high $f_{\rm d}$), the bar is not apparent until $T\approx 2.4$ Gyr.

In order to estimate the bar strength, we use Fourier decomposition of the disc surface density and measure the power in the $m=2$ mode \citep[e.g.][]{Athanassoula2003}. We do this within a specified radius that fully encloses the bar, i.e. $R = 10$ kpc, and divide this radial range in concentric rings of width $\Delta R = 1$ kpc. This approach yields an array of values for the $m=2$ amplitude, one for each radial bin $\Delta R$. The bar strength, generally denoted $A_2$, is a time-dependent quantity variously reported, at each time $T$, as the radially-weighted average of the array of values \citep[e.g.][]{Athanassoula2003}, by the arithmetic average of these values \citep[e.g.][]{gaj17m}, or by the maximum value within the adopted radial range \citep[e.g.][]{kat19x}. Here we follow the latter approach in our estimation of $A_2$, although we find that all methods give comparable results.

The bar strength for both the low-resolution barred galaxy and its high-resolution counterpart are shown in the top left panel of Fig.~\ref{f:a2o}. We find a very consistent behaviour between these simulations, but there are clear differences. In particular, we find that the onset of the bar instability, indicated by a sharp rise in $A_2$, is noticeably delayed on the high resolution simulation, in agreement with \citet{Fujii2018}. Also, the bar appears slightly stronger (i.e. $A_2$ reaches higher values) in the high resolution simulation.

For both low and high resolution, the bar strength rises rapidly up to $A_2 \approx 0.4 - 0.5$ at $T \approx 2.4$ Gyr, after which it flattens out, indicating that the bar is fully formed, and reasonably stable for the duration of the simulation ($T \approx 4.3$ Gyr). The overall rising trend in $A_2$ indicates that the onset of the bar instability is present from the very beginning of the simulation. Remarkably, the same effect is also seen in the {\it smooth} disc (low $f_{\rm d}$) model (Fig.~\ref{f:a2o2}), albeit at much lower amplitude and rate of change with cosmic time. 
Thus, {\it all} HBD models appear to be ultimately unstable to bar distortions \citep[cf.][]{Fujii2018}.

\subsubsection{Bar structural properties} \label{s:bar4}

In the Milky Way, the total stellar mass of the long bar and the bulge structure is $(1.88 \pm 0.12) \times 10^{10}$ \Msun, while the mass of the `inner' disc ($R \lesssim 5$ kpc) is estimated at  $(1.29 \pm 0.12) \times 10^{10}$ \Msun{} \citep{Portail2017}.
In our synthetic galaxy, we have set the total  mass of the classical bulge component at $1.2\times 10^{10}$ \Msun; this does not change during the galaxy's evolution. Furthermore, we measure the composite disc/bar mass at $R\leq 5$ kpc to be
$\sim 2.3 \times 10^{10}$ \Msun\ (at $T\approx 2.4$ Gyr). Thus the total stellar mass within the bar region ($R \leq 5$ kpc) in our model adds up to roughly $3.5 \times 10^{10}$ \Msun, which compares well with the corresponding MW estimate ($3.17 \times 10^{10}$ \Msun).

We measure the total dynamical mass within $r = 2$ kpc (the `bulge region')
to be \mbox{$2.6\times 10^{10}$ \Msun}.
For comparison, \citet{Portail2017} estimate a dynamical mass within the bulge region of \mbox{$(1.85 \pm 0.05) \times 10^{10}$} \Msun\ which translates into a $\sim 17$ percent DM-to-total mass ratio; in our model, we find a corresponding value of $0.3/2.6 \approx 0.11$ or 11 per cent, demonstrating that baryons completely dominate the central regions, both in our model and in the Milky Way.

We estimate the linear size of the bar in our model in the following way: (1) we calculate the iso-density contour at a level of \mbox{$3\times10^5$ \Msun~\kpc$^{-2}$} - guided by the result of Fig. \ref{f:surfdens_bar2} (left); and (2) we record the maximum extension along the major semi-axis defined by this contour. At $T \approx 2.4$ Gyr, we find the bar extends out to roughly 4.5 kpc, in comparison to $\sim$5 kpc of the MW bar. We note that the bar length does not change appreciably by the end of the simulation at $T \approx 4.3$ Gyr.

For convenience, we have summarised all the results just discussed in Tab.~\ref{t:comp}.

\subsubsection{Bar pattern speed and long-term behaviour} \label{s:bar3}


There are several methods to estimate the pattern speed: (1) Fourier decomposition; (2) estimation of Jacobi integral; (3) determining moments of inertia \citep[for a brief overview see][their appendix]{wu18t}; and (4) the Tremaine-Weinberg (TW) method \citep{tre84h}. To the best of our knowledge, only the first three have been applied to simulations as ours \citep[but see][]{zou19w}. One drawback of these approaches is that they do not generally provide a single, well-defined value of the pattern speed. However, the bar in our model has clear figure rotation. This can be best appreciated in the movies provided as supplement material with this paper (see Sec.~\ref{s:data}).

As a result of this ambiguity, one has then to adopt a `definition' of what characterises the pattern speed of the bar. The Fourier method relies on calculating the amplitude ($A_2$) and the phase ($\phi_2$) of the \mbox{$m=2$} mode in a specified radial range (see Sec.~\ref{s:bar2}), and computing the phase change ($\dot{\phi}_2$) for the datum at which $A_2$ reaches a maximum, consistent with the estimate of the bar strength \citep[e.g.][]{kat19x}. Another related approach is to perform a radially-weighted average of $\dot{\phi}_2$ over the specified radial range \citep[e.g.][]{Athanassoula2003}, or simply consider the whole range of values without conveying a special meaning to any in particular \citep[e.g.][]{wu18t}. In all these approaches, the pattern speed is equated to the phase change $\dot{\phi}_2$.

We have applied each of these methods to measure the 'pattern speed' of the bar in our simulation. The results are presented in Fig.~\ref{f:a2o} (bottom left). Interestingly, all three methods show the same trend in the bar slowing down with cosmic time, but with small differences in pattern speed at the outset.

When we attempt to `freeze' the bar using any of the trends thus obtained, none of them appear to track the bar's rotation precisely. Guided by the common behaviour of these trends, we describe the bar pattern speed using a simple function of cosmic time that works remarkably well overall. Our method relies on a visual impression of what defines the long axis of the bar. The eye is very good at averaging over the complex behaviour around and at the ends (ansae) of the bar. We find that the bar's figure rotation is well described by a slow, exponentially declining function of time, $\Omega_p(T) = \Omega_p(0) \exp \left[ -T / \tau \right]$, with $\Omega_p(0) \approx 48.3$ \kms \pkpc and and e-folding time $\tau \approx 20.5$ Gyr.

In addition, we estimated the bar pattern speed using \citet{tre84h}'s method in a selected snapshot of our simulation, corresponding to $T \approx 2.4$ Gyr. To this end, we select all the disc particles within an annulus of inner (outer) radius of 0.5 kpc (10 kpc) and height $\Delta z = 2$ kpc. The result is shown in Fig.~\ref{f:a2o} (top right). The green symbols indicate the measurement perpendicular to the bar major axis; the black symbols, to the measurements parallel to the bar major axis. The former yield a clearer correlation between the mean sightline velocity with radius along the bar.  We include the outcome of the TW analysis in the bottom left panel; it is indicated by a cross. Note that this value is consistent with the actual behaviour of $\Omega_p(T)$ as described by our fit.

In Fig.~\ref{f:a2o} (bottom right), we show a range of recent estimates of the bar pattern speed that partly lie in the dark grey band defined by our temporal model. The width of this band corresponds to a bar age of about 2 Gyr. To accommodate the lower estimates, we add an additional light grey band to extend the time frame by about 3 Gyr. If the onset and evolution of our model is indicative of the Milky Way, the bar first emerged in the last $3-4$ Gyr. This time scale is significantly shorter than a recent estimate of the bar age of $\sim 6$ Gyr from the quenching of the star formation in the bar compared to the inner ring stars \citep{Wylie2021}.

Our fit to the evolution of the pattern speed implies a change rate of $\dot{\Omega}_p \approx -2.3$ \kms \pkpc \pGyr, or roughly 10 percent in 2 Gyr. This deceleration is roughly half the value recently estimated for the bar of the Galaxy from \gaia\ observations \citep{Chiba2021}. Their estimate comes from a resonance drift model, which they use to reproduce the  double-peaked structure (seen in the $U -V$ plane) of the Hercules stream.

Given the above results, we believe that the synthetic bar has global, structural and kinematic properties that compare well with the corresponding properties of the MW's bar. 

\section{Final remarks}

A working model of a barred Milky Way has many practical applications. It can be compared with observations of the Galaxy \citep{robin2012} and with oft-claimed Milky Way analogues emerging from cosmological simulations \citep[e.g.][]{ElBadry2018}.

Our model provides a framework to develop new tests of specific Milky Way properties, e.g., constraints on the bar pattern speed or slowdown rate \citep{Chiba2021}, on the formation of various components \citep[e.g.][]{Loebman2011},
or to understand how the Galaxy responds to internal or external impulses \citep{Bennett2021}.

Our goal here is not to match all known features of the Milky Way. Attempting to achieve such a goal has limited scientific value because it is not grounded in the MW's long and complex formation history.
Instead, we seek to provide a base model to build upon towards a more realistic - and necessarily more complex - dynamical model of the MW, available to the  community.

The present model is defined by roughly 40 free parameters,
of which about half are considered to be of paramount importance to this work. Some of these are constrained by observations, others can be given meaningful values by physical considerations.

Our particular interests are how the model responds to both axisymmetric and non-axisymmetric perturbations, e.g. the presence of a significant gas phase, giant molecular clouds, spiral arms, clumpy dark matter, central star clusters, transitting dwarf galaxies and so forth, all of which shall be addressed in the future.

Our work follows in the wake of \cite{Fujii2018} and \citet{Fujii2019}. Our simulation bridges the gap between these studies in terms of particle resolution, and supports their finding of $f_{\rm d}$ as the key parameter to the construction of an N-body, barred MW surrogate. We confirm that a correct choice of $f_{\rm d}$ appropriate to the Milky Way -- i.e. as constrained by observation -- leads to bar formation within a few billion years. A remarkable revelation is that the emergent bar has properties closely associated with those of the Milky Way once $f_{\rm d}$ is made to match the Galactic value at the outset.

The fact that our results are consistent with \citet{Fujii2019} in every relevant aspect, despite the different approach taken by each team, is encouraging as it suggests that we finally have at our disposal a powerful framework consisting of initialisation codes (GalactICs, \agama) and N-body solvers to construct a realistic, working, self-consistent N-body model for the Milky Way. But we recognise that the current models are still limited because they are incomplete.

Observationally, there are many aspects of the Milky Way's bar that have been addressed in detail, but are not well understood, e.g. the inner X-structure \citep{Combes1990,Wegg2013} seen clearly in side elevation \citep{Ness2016}. Formation mechanisms for box/peanut bulges have a long history \citep{Combes1990,Athanassoula2005,Sellwood2020}. \citet{Quillen2014} showed that bar buckling is not needed for X-formation if the bar slows down, which is consistent with what we observe in our model.

Other features are entirely mysterious, including the long, cold (`superthin') bar that extends beyond the puffed-up inner bar \citep{Hammersley2000}. Such a feature has never been reliably identified in N-body simulations \citep{blandhawthorn2016}. Part of the problem \citep{Sellwood1988} may be confusion arising from the different pattern speeds of the bar and the outer arms, such that these features line up sporadically creating `ansae' at the ends of the bar that lead and trail at different times. We see precisely this behaviour in our simulations.

In the Milky Way, the stellar bar is highly aligned (within a few arcminutes) with the inner disc plane \citep{blandhawthorn2016}, with no evidence for vertical buckling modes that can tilt the bar's long axis \citep[e.g.][]{Khoperskov2019}. The projected, vertical velocity field $V_z$ shows little or no obvious structure such that the bar's action is almost entirely confined to the $x-y$ plane. The vertical buckling signature may have been detected in a handful of external barred galaxies seen face-on \citep{Xiang2021}.
In our models, weak patterns are discernible at the level of $\vert V_z\vert \sim 1$ km s$^{-1}$. This is consistent with our earlier statements that no discernible vertical heating is apparent from the bar's action.

In the language of \citet{Fujii2019}, we have managed to produced a {\em dry} working model of the Milky Way. In the next paper, we examine the inclusion of gas within the \agama\ set up. 
Using the \ramses\ code, we are able to demonstrate long-term equilibrium and a high level of stability in the sense that the gas retains its {\it original} smoothed radial distribution, as specified. To our knowledge, this has not been demonstrated before even though attempts have been made \citep[e.g.][]{Deg2019}. But we demonstrate that the gas fraction, much like the disc to total mass fraction, plays a key role in how the bar forms and evolves subsequently \citep{Athanassoula2013,VillaVargas2010}.

In summary, we present our current working model as a useful framework, but fully recognize that there is no end to the possible refinements that can be considered.

\section{Data availability} \label{s:data}
The full simulation output amounts to roughly 1.8 Tb of compressed data. These as well as movies showing the full evolution of the galaxy are all made publicly available through the website at \url{http://www.physics.usyd.edu.au/\~tepper/proj\_mwbar.html} .

\section*{Acknowledgments}

TTG is funded under JBH's ARC Laureate Fellowship and by generous funding from the Grand Challenge Project on Modern Slavery from the University of Sydney. JBH \& TTG acknowledge the resources and services from the National Computational Infrastructure (NCI), which is supported by the Australian Government. EA acknowledges financial support from the CNES.
EV acknowledges support from STFC via the Consolidated grant to the Institute of Astronomy Cambridge. PM acknowledges funding through a research project grant from the Swedish Research Council (Vetenskapr\aa{}det, Reg: 2017-03721).
We acknowledge both inspiration and insights from our students in the weekly Galacticos meetings, and our colleagues within the GALAH survey team and the ASTRO-3D Centre of Excellence. We are  particularly grateful to Agris Kalnajs for sharing his wisdom  from seven decades of dedicated work on Galactic dynamics.

\bibliographystyle{mnras} 
\bibliography{mw_bar.bib} 

\appendix

\section{Smooth vs. Barred model}

\begin{figure}
\centering
\includegraphics[width=0.45\textwidth]{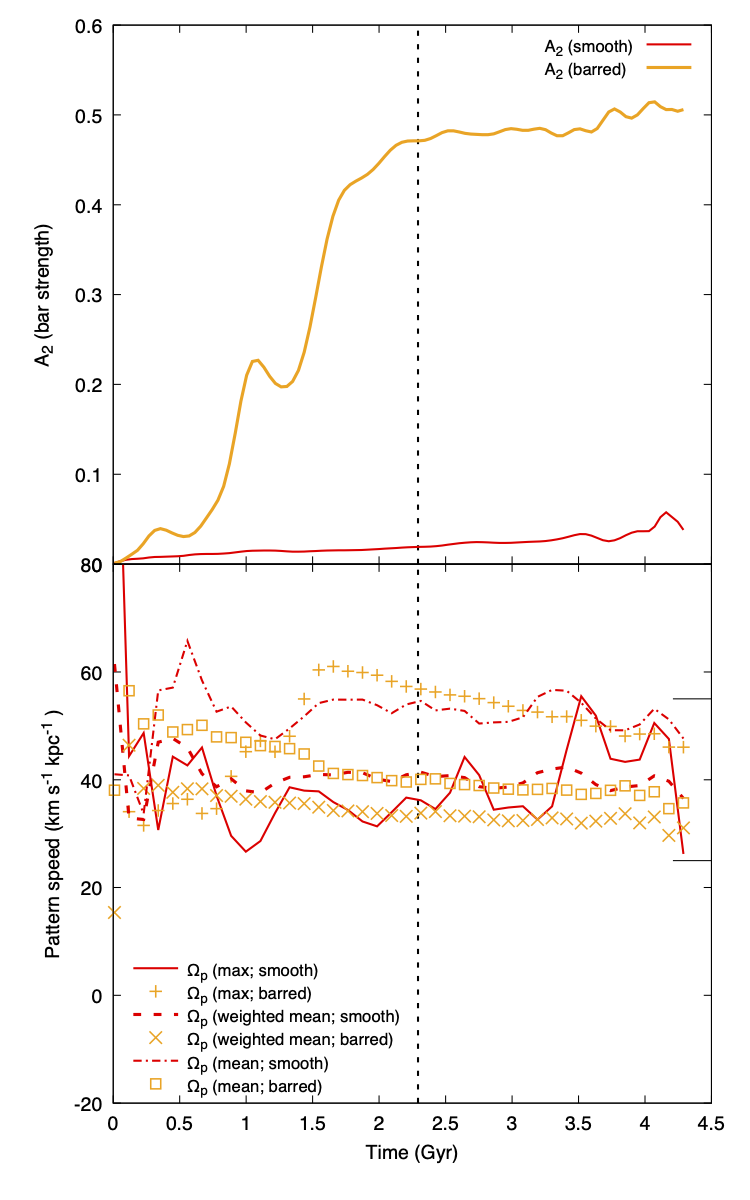}
\vspace{-5pt}
\caption[  ]{ Fourier analysis on our smooth galaxy model (red curves). Note that the orange curve and symbols are identical to those in Fig.~\ref{f:a2o}. The smooth model does not feature a bar, as indicated by the low $A_2$ amplitude, nor does its central region (within 10 kpc) display a well-defined pattern speed. }
\label{f:a2o2}
\end{figure}

In order to verify our method based on Fourier analysis used to measure the bar properties (strength, pattern speed) in our barred model (cf. Fig.~\ref{f:a2o}), we perform the same analysis on our smooth galaxy model. The result is displayed in Fig.~\ref{f:a2o2}. The methodology used here is discussed in detail in Sec. ~\ref{s:bar3}. The main take-away message from this exercise is that the smooth model does not visually develop a bar in roughly $4.3$ Gyr of evolution  \citep[cf.][]{BlandHawthorn2021}, and this is consistent with the low amplitude of $A_2$ over the same time period.

%


\section{Model vs. Observations}

\begin{figure*}
\centering
\includegraphics[width=1.3\textwidth, angle=90]{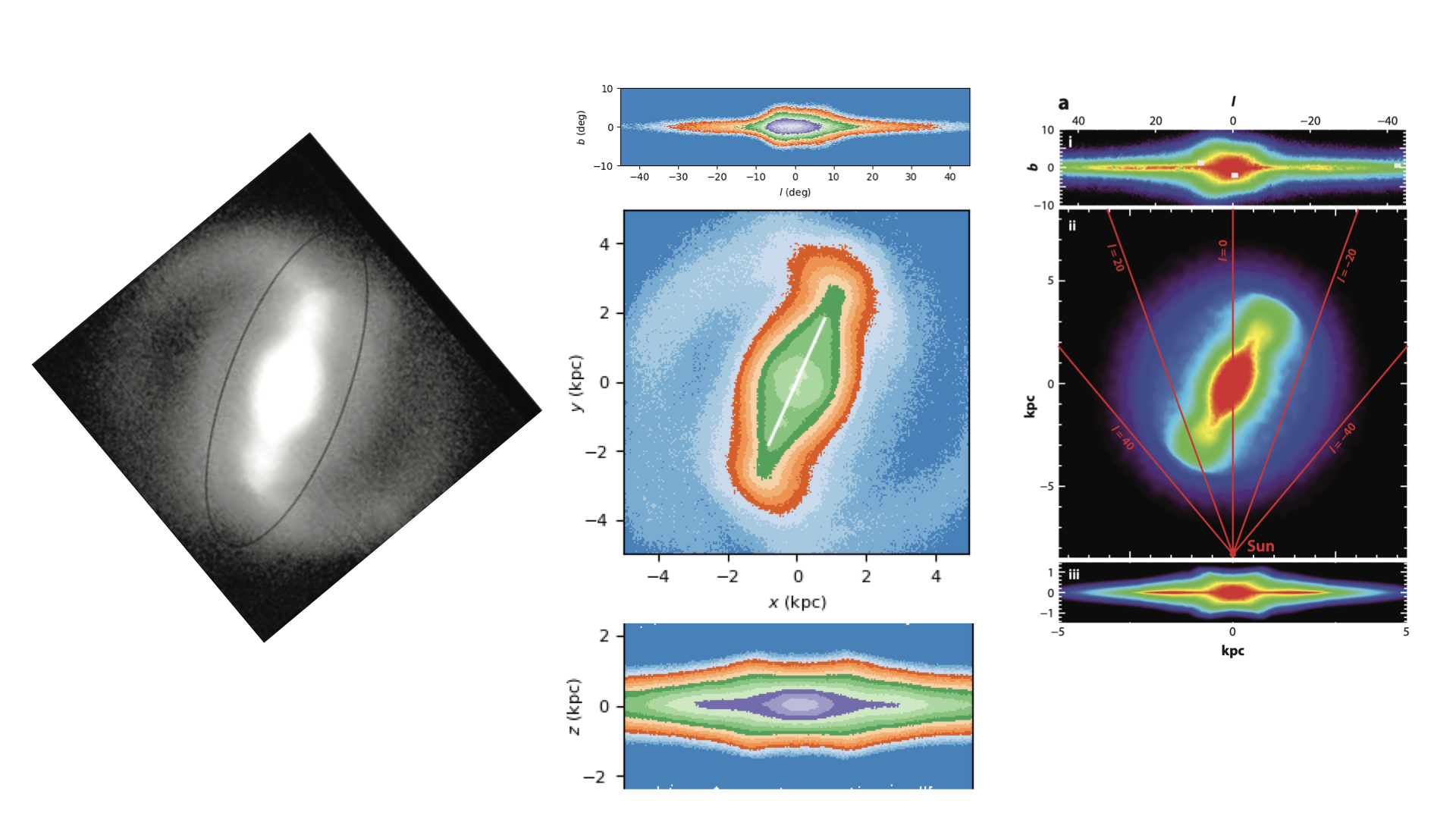}
\vspace{-5pt}
\caption[  ]{ Collage showing a comparison between barred galaxies (MW, NGC4314) and our barred MW model at $T \approx 2.4$ Gyr. Top image adapted from \citet{Wegg2015}; bottom image adapted from \citet{blandhawthorn2016}. }.
\label{f:collage}
\end{figure*}

\begin{figure*}
\centering
\includegraphics[width=1.3\textwidth, angle=90]{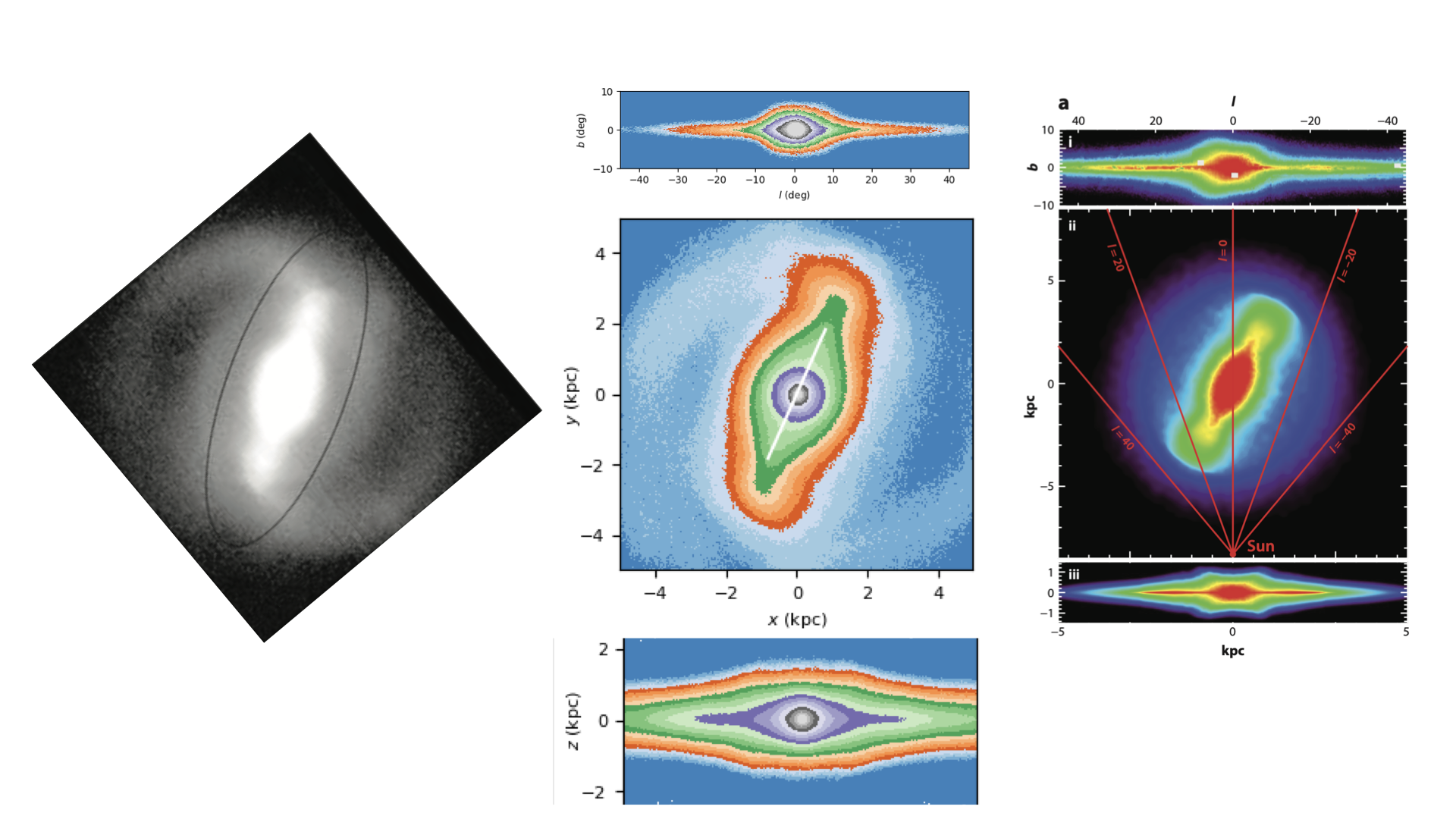}
\vspace{-5pt}
\caption[  ]{ Same as Fig. \ref{f:collage}, but including also the classical bulge in the N-body results (central panels). Top image adapted from \citet{Wegg2015}; bottom image adapted from \citet{blandhawthorn2016} }
\label{f:collage2}
\end{figure*}

In Figs. \ref{f:collage} and \ref{f:collage2} we present a comparison of our synthetic galaxy to data and models of the MW's bar region, and the bar region of an external galaxy, NGC4314. The latter is believed to have face-on morphology similar to the MW \cite[cf.][]{blandhawthorn2016}. The intent of this comparison is to highlight the similarities and differences between our N-body model and the MW in a visual, qualitative manner. 

%
---------------------------------------------
%

\bsp	
\label{lastpage}
\end{document}

%% file: defs_mnras.tex
\definecolor{orange}{rgb}{1.0,0.5,0.}

\DeclareMathOperator{\sech}{sech}

\def\MDM{\ifmmode{\>M_{\textnormal{\sc dm}}}\else{$$M_{\textnormal{\sc dm}}}\fi}

\def\XH{\ifmmode{\>X_{\textnormal{\sc h}}} \else{$X_{\textnormal{\sc h}}$}\fi}
\def\nH{\ifmmode{\>n_{\textnormal{\sc h}}} \else{$n_{\textnormal{\sc h}}$}\fi}

\def\maspyr{\ifmmode{\>\textnormal{mas~yr}^{-1}}\else{mas~yr$^{-1}$}\fi}

\def\mG{\ifmmode{\>\mu\mathrm{G}}\else{$\mu$G}\fi}
\def\erg{\ifmmode{\> {\rm erg}}\else{erg}\fi}
\def\keV{\ifmmode{\> {\rm keV}}\else{keV}\fi}

\def\deg{\ifmmode{\>^{\circ}}\else{$^{\circ}$}\fi}
\def\onedeg{\ifmmode{\>1^{\circ}}\else{$1^{\circ}$}\fi}

\def\xvir{\ifmmode{\>\!x_{vir}}\else{$x_{vir}$}\fi}
\def\Mvir{\ifmmode{\>\!M_{vir} }\else{$M_{vir} $}\fi}
\def\rvir{\ifmmode{\>\!r_{vir}}\else{$r_{vir}$}\fi}
\def\vvir{\ifmmode{\>\!v_{vir}}\else{$v_{vir}$}\fi}
\def\Vvir{\ifmmode{\>\!V_{vir} }\else{$V_{vir} $}\fi}

\def\tratio{\ifmmode{\>\tau}\else{$\tau$}\fi}

\def\rms{\ifmmode{\>r_{\textnormal{\sc ms}}}\else{$r_{\textnormal{\sc ms}}$}\fi}

\def\Mpc{\ifmmode{\>\!{\rm Mpc}} \else{Mpc}\fi}

\def\kpc{\ifmmode{\>\!{\rm kpc}} \else{kpc}\fi}
\def\pkpc{\ifmmode{\>\!{\rm kpc}^{-1}}\else{kpc$^{-1}$} \fi}

\def\pc{\ifmmode{\>\!{\rm pc}} \else{pc}\fi}

\def\Gyr{\ifmmode{\>\!{\rm Gyr}} \else{Gyr}\fi}
\def\pGyr{\ifmmode{\>\!{\rm Gyr}^{-1}} \else{Gyr}$^{-1}$\fi}

\def\Myr{\ifmmode{\>\!{\rm Myr}} \else{Myr}\fi}

\def\yr{\ifmmode{\>\!{\rm yr}} \else{yr}\fi}
\def\pyr{\ifmmode{\>\!{\rm yr}^{-1}}\else{yr$^{-1}$} \fi}

\def\s{\ifmmode{\>\!{\rm s}}\else{s}\fi}
\def\ps{\ifmmode{\>\!{\rm s}^{-1}}\else{s$^{-1}$}\fi}
\def\Hz{\ifmmode{\>\!{\rm Hz}}\else{Hz}\fi}

\def\kms{\ifmmode{\>\!{\rm km\,s}^{-1}}\else{km~s$^{-1}$}\fi}

\def\K{\ifmmode{\>\!{\rm K}}\else{K}\fi}

\def\sr{\ifmmode{\>\!{\rm sr}}\else{sr}\fi}
\def\psr{\ifmmode{\>\!{\rm sr}^{-1}}\else{sr$^{-1}$}\fi}
\def\arcs{\ifmmode{\>\!{\rm arcsec}}\else{arcsec}\fi}
\def\parcs{\ifmmode{\>\!{\rm arcsec}^{-1}}\else{arcsec${-1}$}\fi}
\def\parcss{\ifmmode{\>\!{\rm arcsec}^{-2}}\else{arcsec${-2}$}\fi}

\def\cm{\ifmmode{\>\!{\rm cm}}\else{cm}\fi}
\def\cc{\ifmmode{\>\!{\rm cm}^{3}}\else{cm$^{3}$}\fi}
\def\sqc{\ifmmode{\>\!{\rm cm}^{2}}\else{cm$^{2}$}\fi}
\def\pcc{\ifmmode{\>\!{\rm cm}^{-3}}\else{cm$^{-3}$}\fi}
\def\psc{\ifmmode{\>\!{\rm cm}^{-2}}\else{cm$^{-2}$}\fi}

\def\g{\ifmmode{\>\!{\rm g}}\else{g}\fi}
\def\Msun{\ifmmode{\>\!{\rm M}_{\odot}}\else{M$_{\odot}$}\fi}
\def\hMsun{\ifmmode{\> h^{-1}{\rm M}_{\odot}}\else{$h^{-1}$M$_{\odot}$}\fi}

\def\Zsun{\ifmmode{\>\!{\rm Z}_{\odot}}\else{Z$_{\odot}$}\fi}

\def\Lsun{\ifmmode{\>\!{\rm L}_{\odot}}\else{L$_{\odot}$}\fi}

\def\rayl{\ifmmode{\>\!{\rm R}}\else{R}\fi}
\def\mR{\ifmmode{\>\!{\rm mR}}\else{mR}\fi}

\renewcommand{\ion}[2]{\hbox{#1\,{\sc #2}}}

\def\lya{\ifmmode{\>\!{\rm Ly}\alpha}\else{Ly$\alpha$}\fi}

\def\Ha{\ifmmode{\>\!{\rm H}\alpha}\else{H$\alpha$}\fi}
\def\Hb{\ifmmode{\>\!{\rm H}\beta}\else{H$\beta$}\fi}

\def\HI{\ifmmode{\> \textnormal{\ion{H}{i}}} \else{\ion{H}{i}}\fi}
\def\HII{\ifmmode{\> \textnormal{\ion{H}{ii}}} \else{\ion{H}{ii}}\fi}
\def\CIV{\ifmmode{\> \textnormal{\ion{C}{iv}}} \else{\ion{C}{iv}}\fi}
\def\SiIV{\ifmmode{\> \textnormal{\ion{S}{iv}}} \else{\ion{Si}{iv}}\fi}

\def\NH{\ifmmode{\> {\rm N}_{\rm H}} \else{N$_{\rm H}$}\fi}
\def\Ng{\ifmmode{\> {\rm N}_{\rm gas}} \else{N$_{\rm gas}$}\fi}
\def\NHI{\ifmmode{\> {\rm N}_{\HI}} \else{N$_{\HI}$}\fi}
\def\MHI{\ifmmode{\> {\rm M}_{ \HI}} \else{M$_{\HI}$}\fi}

\def\mua{\ifmmode{\>\mu_{ \textnormal{\Ha}}}\else{$\mu_{ \textnormal{\Ha}}$}\fi}
\def\alphabha{\ifmmode{\>\alpha_{B}^{(\textnormal{\Ha})}}\else{$\alpha_{B}^{(\textnormal{\Ha})}$}\fi}